\documentclass{iopart}
\pdfoutput=1
\usepackage{graphicx}
\usepackage{xspace}
\newcommand{\Al}{$^{26}$Al\xspace}

\newcommand{\Msol}{M\ensuremath{_\odot}\xspace}

\begin{document}
\title{New insights from cosmic gamma rays}

\author{Roland Diehl}

\address{Max Planck Institut f\"ur extraterrestrische Physik, D-85748 Garching, Germany}

\ead{rod@mpe.mpg.de}

\begin{abstract}
The measurement of gamma rays from cosmic sources at $\sim$MeV energies is one of the key tools for nuclear astrophysics, in its study of nuclear reactions and their impacts on objects and phenomena throughout the universe. Gamma rays trace nuclear processes most directly, as they originate from nuclear transitions following radioactive decays or high-energy collisions with excitation of nuclei. Additionally, the unique gamma-ray signature from the annihilation of positrons falls into this astronomical window and is discussed here: Cosmic positrons are often produced from $\beta$-decays, thus also of nuclear physics origins. The nuclear reactions leading to radioactive isotopes occur inside stars and stellar explosions, which therefore constitute the main objects of such studies. In recent years, both thermonuclear and core-collapse supernova radioactivities have been measured though $^{56}$Ni, $^{56}$Co, and $^{44}$Ti lines, and a beginning has thus been made to complement conventional supernova observations with such measurements of the prime energy sources of supernova light created in their deep interiors. The diffuse radioactive afterglow of massive-star nucleosynthesis in gamma rays is now being exploited towards astrophysical studies on how massive stars feed back their energy and ejecta into interstellar gas, as part of the cosmic cycle of matter through generations of stars enriching the interstellar gas and stars with metals. Large interstellar cavities and superbubbles have been recognised to be the dominating structures where new massive-star ejecta are injected, from $^{26}$Al gamma-ray spectroscopy. Also, constraints on the complex interiors of stars derive from the ratio of $^{60}$Fe/$^{26}$Al gamma rays. Finally, the puzzling bulge-dominated intensity distribution of positron annihilation gamma rays is measured in greater detail, but still not understood; a recent microquasar flare provided evidence that such objects may be prime sources for positrons in interstellar space, rather than distributed nucleosynthesis. We also briefly discuss the status and prospects for astronomy with telescopes for the nuclear-radiation energy window. 
\end{abstract}

\section{Introduction}
The nuclear processes in cosmic objects produce, among others, unstable, radioactive isotopes (from nucleosynthesis), or nuclei excited above their ground states (from energetic collisions). Therefore, a measurement of cosmic gamma rays makes uses of electromagnetic radiation (i.e. \emph{astronomy}) as a remote probe to learn about such nuclear processes in distant sources. This complements studies of presolar grains, which are being found as inclusions in meteorites, and provide a material messenger from cosmic nuclear reactions from sources outside the solar system. Together, these are the key tools for nuclear astrophysics for the study of nuclear reactions and their impacts on objects and phenomena throughout the universe. All other studies rely on less direct consequences of nuclear reactions: Supernova light and spectra, for example, originate from the photosphere of a supernova, which is rather remote from the inner energy source, and complex radiation transport links the observables to their physical origins. Nevertheless, astronomical precision is more developed at wavelengths longer than those of this gamma-ray range, when astronomical instrumentation is capable to collect and focus radiation that is less penetrating than gamma-rays. 

This paper may be seen as a complement to a similar report written ten years ago from a corresponding lecture series at the Santa Tecla school \cite{Diehl:2007a}, and to a recent review article \cite{Diehl:2013}. Further background material may be found in a recent textbook \cite{Diehl:2011b}.

\begin{figure}%[th] %%%%%%%%%%%%%%%%%%%%%%%%%%%%%%%%%%
\centering
\includegraphics[width=1.0\textwidth]{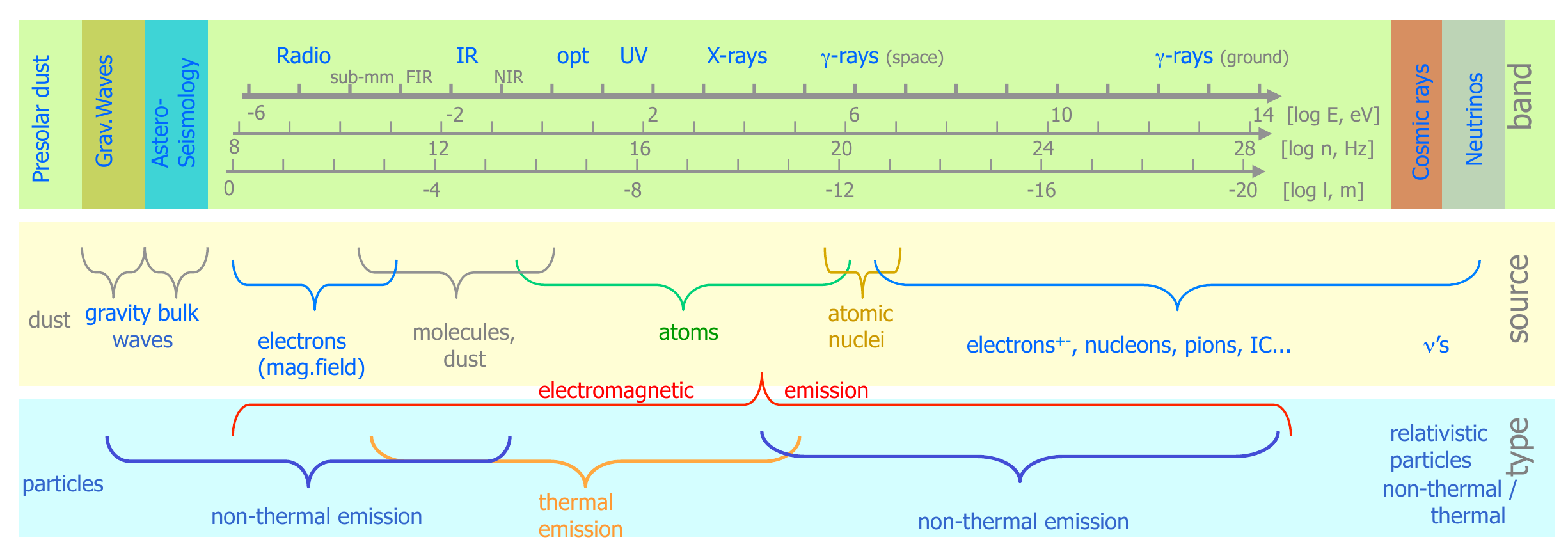}
\caption{The electromagnetic spectrum covers about 20 decades of photon energy. Other astronomical messengers are listed on the sides. The different physical processes per energy are indicated below. Nuclear processes are reflected by photons in the two decades of 0.1 to 10~MeV only.}
\label{fig_em-spectrum}
\end{figure}%%%%%%%%%%%%%%%%%%%%%%%%%%%%%%%%%%

\section{Nuclear astronomy in the context of astrophysics messengers}
Measurements of nuclear radiation and its properties from cosmic sources is difficult. After pioneering detectors on the Apollo spacecraft and in the context of first explorations of the near earth space and solar radiation, it had become clear that radiation at nuclear energies was an observational fact, and that it presented new astronomical constraints \cite{Trombka:1978,Trombka:1983}, such as the \emph{MeV bump} of the diffuse gamma-ray background, line signatures in the high-energy emission of solar flares, and gamma ray bursts. The first Galaxy-scale sources seen in such gamma rays were the lines attributed to positron annihilation reported after a balloon experiment from the central Galaxy in 1964 and the line from  \Al radioactivity gamma-rays also from the central Galaxy region reported from the 1978/79 measurements with the HEAO-C satellite \cite{Mahoney:1982}. But it took many years to further advance the techniques of gamma-ray telescopes, mainly due to the penetrating nature of the gamma rays, which prevent photon collection and concentration by mirrors or lenses, and due to activation by cosmic rays of the instrument materials to generate a large instrumental background  \cite{Bertsch:1988}. Therefore, even after two major observatory missions for imaging and spectroscopy \cite{Gehrels:1993,Schoenfelder:1996,Winkler:2003,Vedrenne:2003,Diehl:2013}, currently achieved sensitivities allow to only see Galactic sources and exceptionally-bright supernovae from up to few Mpc distance, and spatial resolution is of order 2-3 degrees only. The spectral resolution, however, has been advanced to and beyond astrophysical needs and expectations, allowing line shape constraints in detail for the lines and sources of interest \cite{Diehl:2013b}.

\begin{figure}%[th] %%%%%%%%%%%%%%%%%%%%%%%%%%%%%%%%%%
\centering
\includegraphics[width=1.0\textwidth]{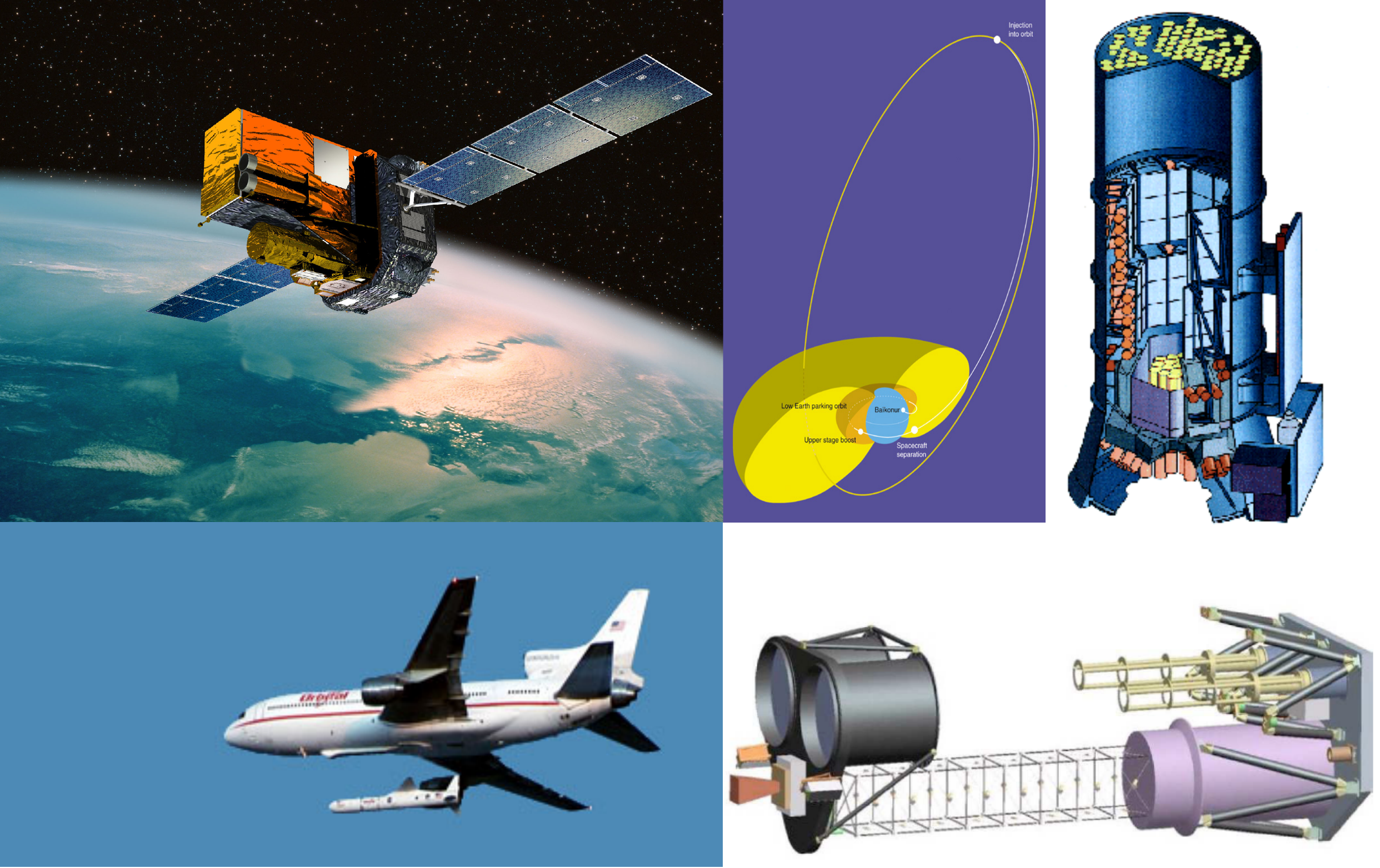} 
\caption{The current instrumentation measuring cosmic photons in the nuclear radiation window. ESA's INTEGRAL mission \cite{Winkler:2003} was launched in 2002 and is currently in its extended mission phase after its 5-year nominal mission \emph{top}; also its excentric orbit is shown \emph{middle} to minimise background variations. The \emph{righthand} figure shows the spectrometer SPI. The NuSTAR X-ray telescope \cite{Harrison:2013} with a focusing mirror on a boom (\emph{bottom}) operates at energies up to 80~keV and was launched from an air plane with a Pegasus rocket in 2012, and also is in its extended phase after its nominal 2-year mission.}
\label{fig_missions}
\end{figure}%%%%%%%%%%%%%%%%%%%%%%%%%%%%%%%%%%

The nuclear window of astronomy (see Fig.\ref{fig_em-spectrum}) may be defined between the lowest-energy nuclear transitions (of $^{44}$Ti at 68 and 78 keV, and $^{60}$Fe at 58~keV) and the highest energies of such transitions from cosmic sources of $\sim$15~MeV (more specifically, the $^{16}$O transitions at 6.1 and 7 MeV and the $^{12}$C transition at 15.1 MeV). This window also includes the signatures of the annihilation of positrons with its prominent line at 511~keV, which, strictly speaking, is not of \emph{nuclear} origin, but unique in its astrophysical information about high-energy sources and processes. Exploration of this astronomical window is fairly recent, starting $\sim$50 years ago \cite{Lingenfelter:1978}. INTEGRAL for the first time combines adequate spectral resolution for line identification and line shape measurements with all-sky exposure.

Other information about such high-energy processes of interest are also not easy to obtain: Meteoritic inclusions have been found to contain a wealth of isotopic information \cite{Clayton:2004}. This derives from measurements of material probes in the laboratory with exquisite precision. The results can be related to sources of nucleosynthesis, yet the formation of presolar dust grains and the journey of these through meteorites to our terrestrial laboratory leaves some systematic issues and uncertainties that limit such interpretation. X-ray spectroscopy also addresses high energy physics and abundances of nuclei in hot gas near nucleosynthesis events. But here potential confusion with atomic re-formation processes and the corresponding astrophysical processes, as well as underlying thermal emission, make it difficult to extract nucleosynthesis information. 
Penetration to dense interiors of cosmic sources of interest is partially possible with gamma rays, more than at X-rays or even lower energies, but still severely affected by radiation re-processing e.g. in stars and supernovae, in particular in comparison with neutrinos. This has been demonstrated in recent years with detailed spectroscopy of solar neutrinos \cite{BOREXINO-Collaboration:2008,BOREXINO-Collaboration:2014}; but the Sun may remain the only source for which neutrino spectroscopy at nuclear physics detail is feasible. Neutrino and gravitational-wave signatures are among the hopes of astrophysicists to make use of new messengers in order to learn more about the interiors of stars, stellar collisions, and explosions. 
But learning about the characteristic signatures in new observational windows, and also the biases and systematic effect, is tedious and takes decades, typically.  The astronomical window of nuclear radiation is now explored since about 2--3 decades; in spite of its restrictions, it will remain worth exploring through improved instrumentation.

\section{Science quests and lessons for nuclear astronomy}
Nuclear-physics processes in cosmic sites are mainly the nuclear reactions in sources of cosmic nucleosynthesis. Inner regions in stars, their core in all cases, and shells for stars beyond the main sequence phase of their initial evolution, but also stellar explosions are prime objects for science investigations using nuclear observations. Furthermore, cosmic sites where high-energy reactions occur or which are cosmic extremes of gravity or electromagnetism have become prominent objects of astrophysical studies also making use of nuclear emission processes.

Interiors of stars are opaque and can be studied only indirectly; but in particular their later evolution beyond core hydrogen burning is plagued with substantial uncertainties \cite{Jones:2015,Pignatari:2013} from complex hydrostatical equilibrium rearrangements that involve material transport with intrinsic 3-dimensionality such as convection and radiation transport through materials of varying densities and composition. The consequences of this late stellar evolution are significant, because late interior evolution synthesises many nuclei which contribute to cosmic metal enrichment, e.g. from the s-process in giant stars whose products are ejected through winds and major contaminations of dust grains found in stardust inclusions of meteorites \cite{Pignatari:2010,Pignatari:2013a}. Also, the core sizes of later pre-supernova stages are determined in this late evolution, as are the carbon to oxygen ratio. Not knowing all this interior complexity, models of stellar evolution carry systematic uncertainties which a.o. result in uncertainties of lifetimes till core collapse supernova or strong wind phases.   

\subsection{Supernovae}
The bright display of supernovae, rising within a few days to a maximum luminosity that rivals galaxies, and falling of slowly at a time scale of months, is due to energy input from radioactive decay \cite{Clayton:1969}. The $^{56}$Ni produced in the supernova decays through gamma-ray and positron emission, and heats the supernova envelope to provide such a spectacular astronomical phenomenon. 

\begin{figure}%[th] %%%%%%%%%%%%%%%%%%%%%%%%%%%%%%%%%%
\centering
\includegraphics[width=0.6\textwidth]{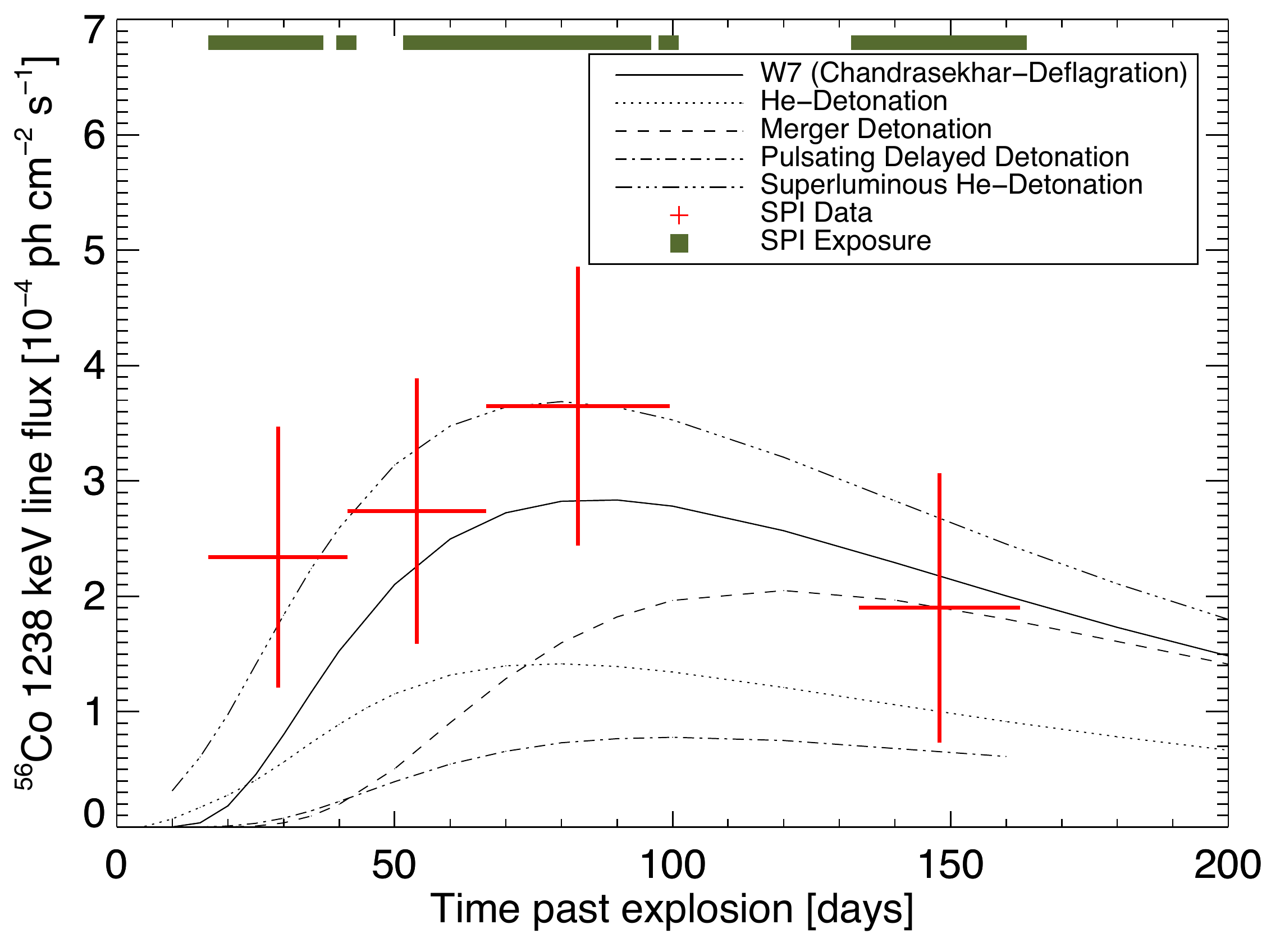} 
\caption{The escape of gamma rays from $^{56}$Co decay in a SN Ia is expected to peak only 60--120 days after the explosion, while the optical brightness maximum occurs typically only 20 days past explosion. Here, several model variants are shown (\emph{continuous and dashed lines})\cite{The:2014}, together with data points from SN2014J measured with INTEGRAL  for the line at 847 keV \cite{Diehl:2015}.}
\label{fig_SN2014J_847LC}
\end{figure}%%%%%%%%%%%%%%%%%%%%%%%%%%%%%%%%%%

Observing gamma rays from the primary energy source (see Fig.~\ref{fig_SN2014J_847LC}) has always been a goal, as well as a dream, for gamma-ray astronomy \cite{Clayton:1969}. The brightest such sources occur quite rare in our nearby universe, and are supernovae of type Ia which produce about 0.5~\Msol and thus about ten times as much $^{56}$Ni as core-collapse supernovae do. In January 2014, luckily one such event occurred at just 3.5~Mpc distance \cite{Fossey:2014aa}, in the nearby star-forming galaxy M82. 

\begin{figure}%[th] %%%%%%%%%%%%%%%%%%%%%%%%%%%%%%%%%%
\centering
\includegraphics[width=0.45\textwidth]{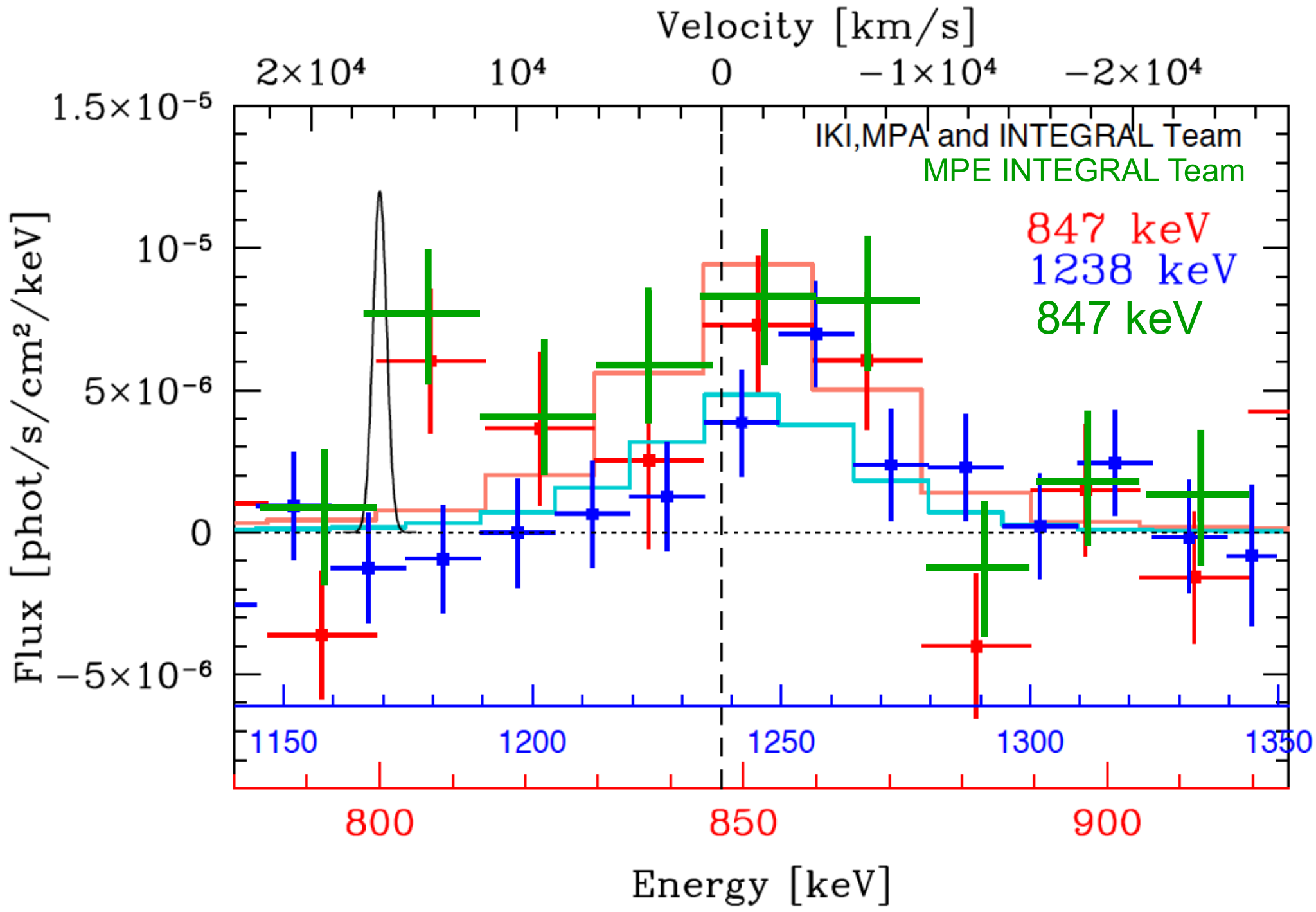} 
\includegraphics[width=0.5\textwidth]{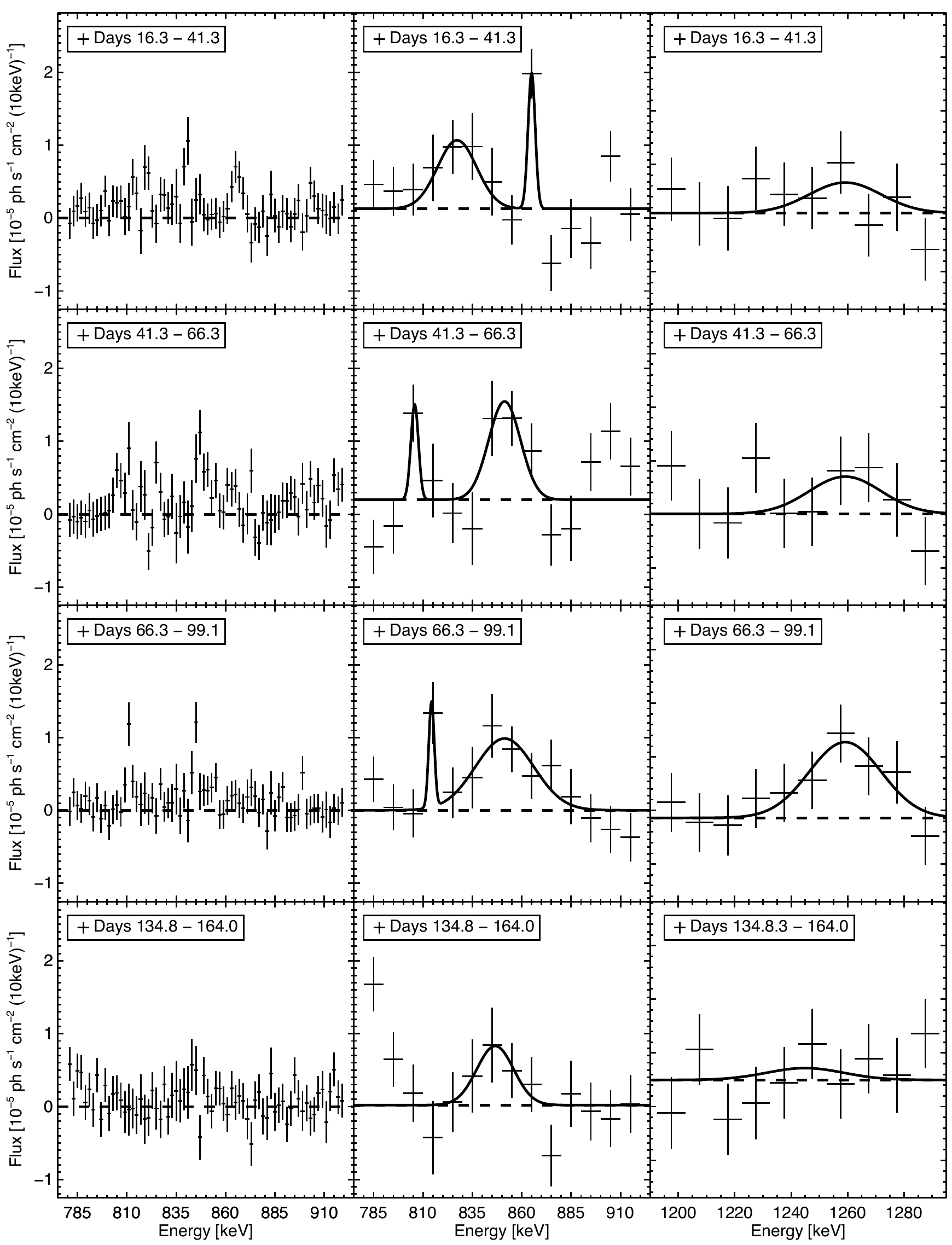} 
\caption{The 847 keV line from $^{56}$Co decay from SN2014J measured with INTEGRAL/SPI  appears broad, and consistently found from two different teams if analysed in broad energy bins \cite{Churazov:2014a,Diehl:2015} (\emph{left}).
The line intensity  appears to rise irregularly, if analysed in fine and broad energy bins and for different phases during the increasing transparency of the supernova envelope \cite{Diehl:2015} (\emph{right}).}
\label{fig_SN2014J_847shape}
\end{figure}

INTEGRAL invested almost half a year of observing time to follow the expected brightening of this supernova in $^{56}$Co decay gamma-rays, which are expected to leak out of the supernova after several weeks towards a maximum brightness expected about three months after the explosion. From the observations, indeed, Doppler broadened lines at 847 and 1238~keV were measured \cite{Churazov:2014a,Diehl:2015} (see Fig.~\ref{fig_SN2014J_847shape}). This is a great triumph of supernova science, and confirms our general understanding of a white dwarf being disrupted by Carbon fusion to power the explosion. The brightness of the lines corresponds to about 0.5~\Msol of $^{56}$Ni produced, as expected. 

There was a surprise, however: INTEGRAL started to observe the supernova 17 days after the inferred explosion date of Jan 14.75 2014 (UT). This seemed too late to test for presence of $^{56}$Ni on the surface, as would be expected by the somewhat exotic \emph{He cap} model variant of SN Ia, considering the decay time of $^{56}$Ni of 8.8~days. But INTEGRAL data provided a clear indication of the presence of the characteristic $^{56}$Ni decay lines at 158 and 812 keV energy \cite{Diehl:2014}. This was unexpected, and points to a rather significant mass of $^{56}$Ni near the surface, as much as 10\% of the total. Such constraints are important to find out what the progenitors of supernovae of type Ia are, and how the explosion actually occurs \cite{Hillebrandt:2013aa}. Therefore, it was hypothesised that the explosion might have been triggered by a surface helium explosion from a helium belt that has built up rather rapidly from matter accreted from the companion star, and thus that the companion also is rather evolved and sheds helium, rather than having a hydrogen envelope\cite{Diehl:2014}. This makes SN2014J appear as an intermediate between a standard white-dwarf / main-sequence binary and a double degenerate system of two white dwarfs. The amount of $^{56}$Ni and other constraints, however, suggest that the mass of the disrupted star was rather high and near the Chandrasekhar limit. The narrowness of the observed gamma-ray lines then suggests that our aspect of the SN2014J progenitor binary system is rather face-on than edge-on. 

%%%%%%%%%%%%%%%%%%%%%%%%%%%%%%%%%%

In many ways this supernova appears to be rather \emph{standard}, but nevertheless shows signs of special surface action early on \cite{Goobar:2014aa}. That may teach us that the \emph{standard} explosions of this type may \emph{appear} as rather spherically symmetric unfoldings of the debris (which re-radiates the radioactive energy input, and may thus have a smoothing effect as a bolometer); but the explosion itself does not necessarily also have to reflect the same spherically symmetric properties, and may in fact allow more complex ignition processes to be involved \cite{Ropke:2011aa}. 

In a way, this is also hinted at by high resolution spectroscopy of the emerging $^{56}$Co decay lines (see Fig.~\ref{fig_SN2014J_847shape}):  The appearance is less smooth and gradual than would be expected by a smooth unfolding of a spherically symmetric envelope including the inner $^{56}$Ni ashes  \cite{Diehl:2015}. Gamma rays at energies around the $^{56}$Co laboratory values appear to come at different times, to gradually build up the Doppler broadened ~40 keV wide lines seen at brightness maximum. But the light curve rises more slowly than standard models suggest, and indicate irregularity that points to a heterogeneous distribution of $^{56}$Ni throughout the expanding supernova. Thus, while being an impressive confirmation of our general understanding of thermonuclear supernovae, the gamma-ray data also present some interesting hints and puzzles upon a closer look. Similar to the SN1987A event, which changed how we thought core-collapse supernovae to arise, SN2014J has the potential to lead to major refinements and changes to our supernova type Ia models.

\begin{figure}%[th] %%%%%%%%%%%%%%%%%%%%%%%%%%%%%%%%%%
\centering
\includegraphics[width=0.5\textwidth]{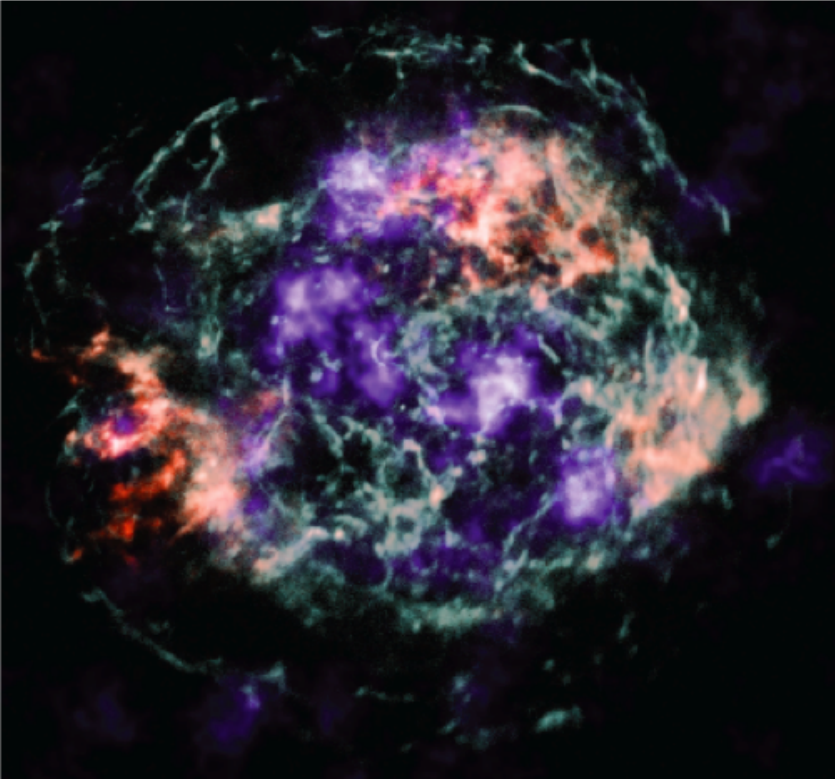} 
\caption{The $^{44}$Ti decay image of Cas A, as measured with NuSTAR from the hard X-ray lines of $^{44}$Sc de-excitation. This image resolves the emission regions, and shows irregular and clumpy ejecta (from \cite{Grefenstette:2014})}
\label{fig_CasA_image}
\end{figure}%%%%%%%%%%%%%%%%%%%%%%%%%%%%%%%%%%

With respect to core collapse supernovae, SN1987A has been teaching us new aspects ever since the explosion almost thirty years ago, as we can follow the phenomena which arise and evolve as we are able to observe them at just 50 kpc distance. Recently, the NuSTAR hard X-ray mirror telescope has been able to measure the low-energy line at 68~keV from $^{44}$Ti decay with great precision \cite{Boggs:2015}. This confirms the earlier INTEGRAL detection of these lines \cite{Grebenev:2012} and thus the direct proof of $^{44}$Ti decay that had been inferred indirectly only from the late, shallow, decline of the bolometric emission of the young supernova remnant. The NuSTAR measurement reports a lower intensity than what INTEGRAL had reported, and the 1.5~10$^{-4}$~\Msol of $^{44}$Ti which are inferred are still higher than standard model yields, but consistent with the late-lightcurve inferred values. The slight red-shift of the $^{44}$Ti line also confirms other indications that the inner explosion had some deviations from spherical symmetry.

The NuSTAR image in $^{44}$Ti 68~keV emission \cite{Grefenstette:2014} (Fig.~\ref{fig_CasA_image}) is a great achievement for nuclear astrophysics. It clearly demonstrates that such deviations from spherically-symmetric explosions, and clumpy structures in the inner ejecta, are an observational fact. Moreover, the fact that the Chandra X-ray image showing Fe as seen in recombination lines differs in detail from the NuSTAR image also is interesting: Explosion physics strongly suggests that Ti and Fe both arise from th very inner supernova regions, and thus should be co-located; it is difficult to imagine a process that would be able to separate them as much as these image comparisons suggest. Rather, it seems plausible that the X-ray recombination emission includes a bias from the ionisation of Fe, which thus does not reflect \emph{all of} the Fe in the expanding remnant. The Fe that is co-located with the $^{44}$Ti emission clumps apparently is not X-ray bright on average. 

\begin{figure}%[th] %%%%%%%%%%%%%%%%%%%%%%%%%%%%%%%%%%
\centering
\includegraphics[width=0.3\textwidth]{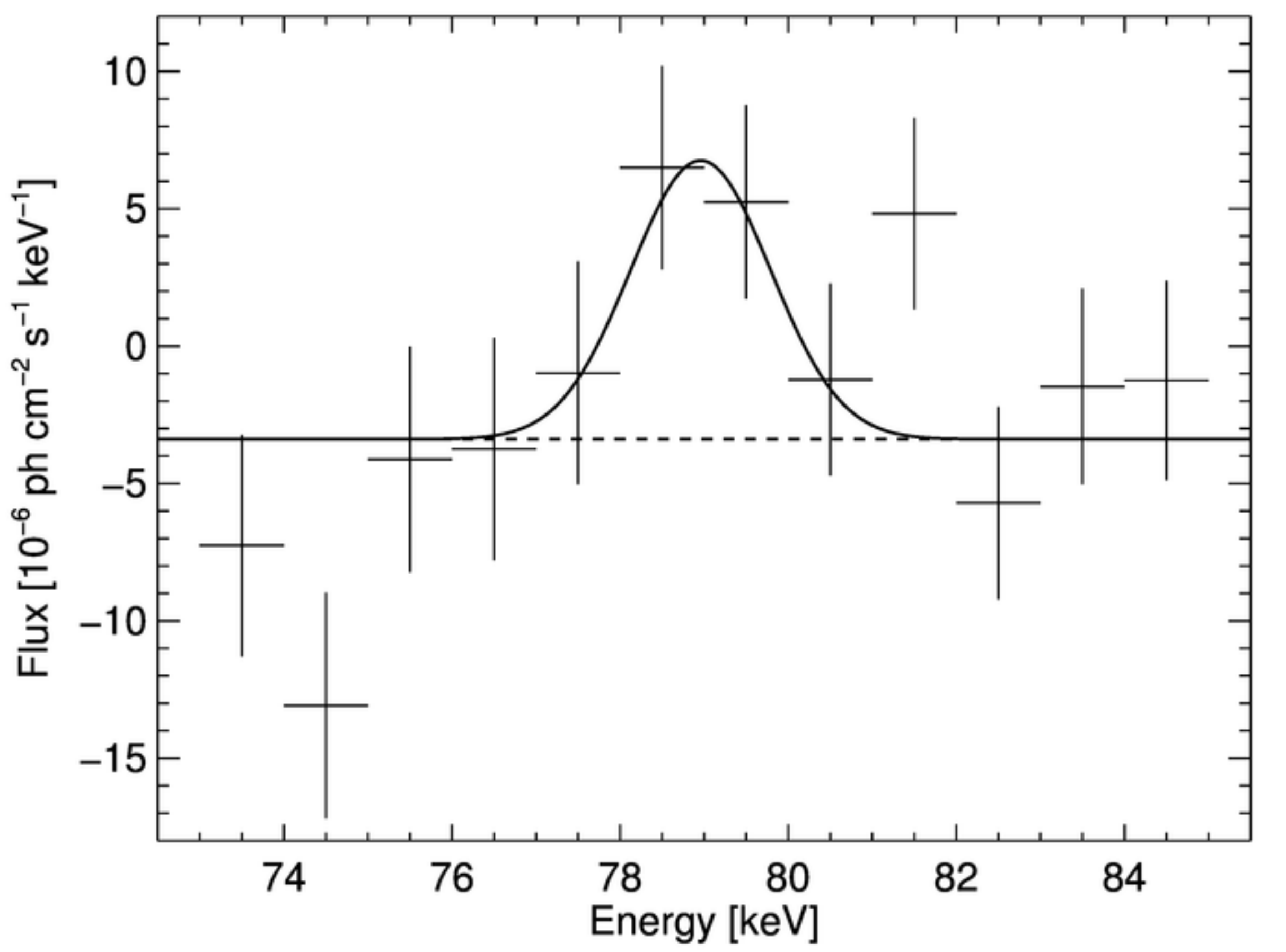} 
\includegraphics[width=0.3\textwidth]{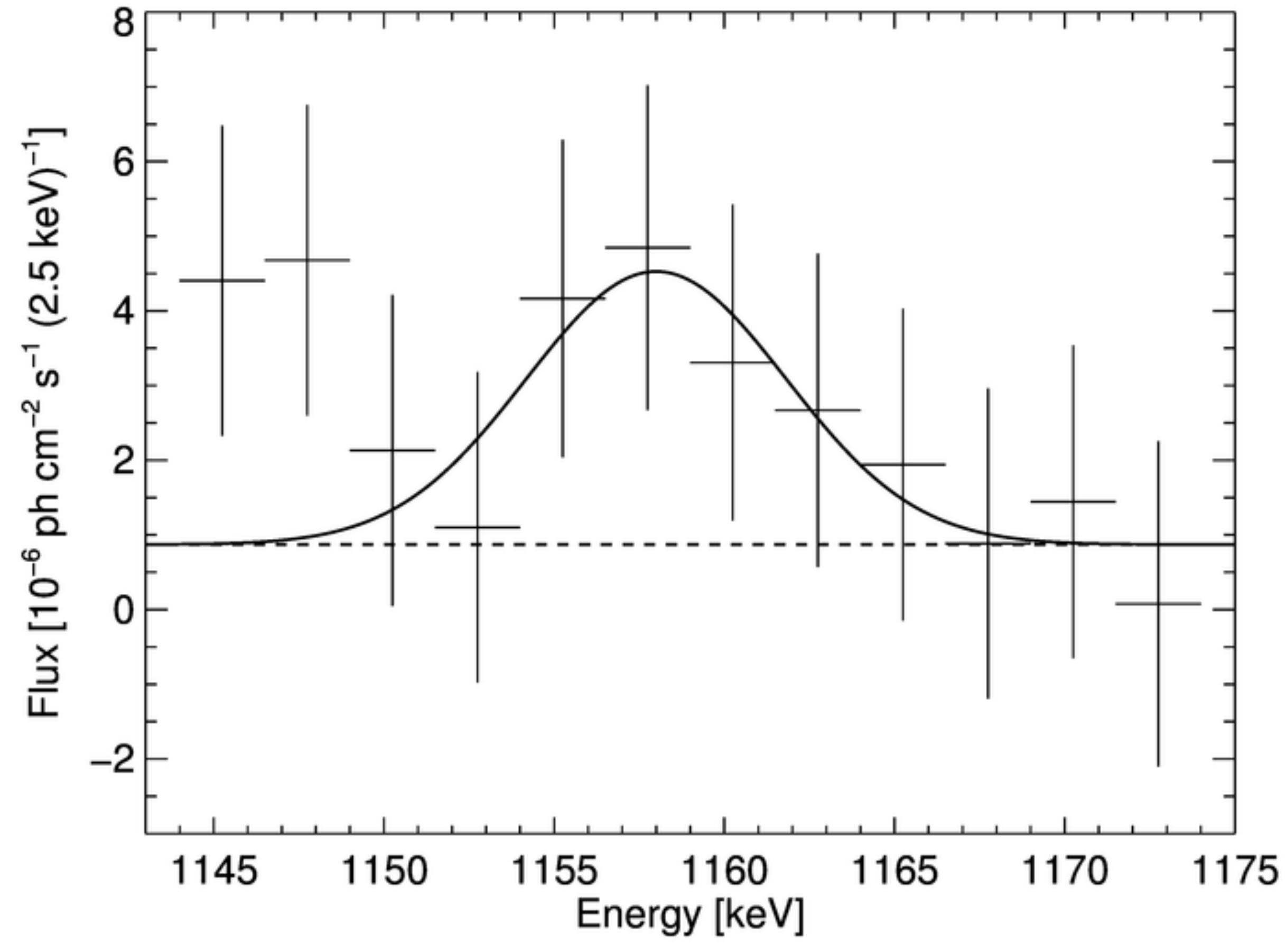} 
\caption{The lines from $^{44}$Ti decay from Cas A, from de-excitation of $^{44}$Sc (\emph{left}), and from de-excitation of $^{44}$Ca (\emph{right}) as measured with SPI/INTEGRAL. (from \cite{Siegert:2015})}
\label{fig_CasA_linespectra}
\end{figure}%%%%%%%%%%%%%%%%%%%%%%%%%%%%%%%%%%

\begin{figure}%[th] %%%%%%%%%%%%%%%%%%%%%%%%%%%%%%%%%%
\centering
\includegraphics[width=0.6\textwidth]{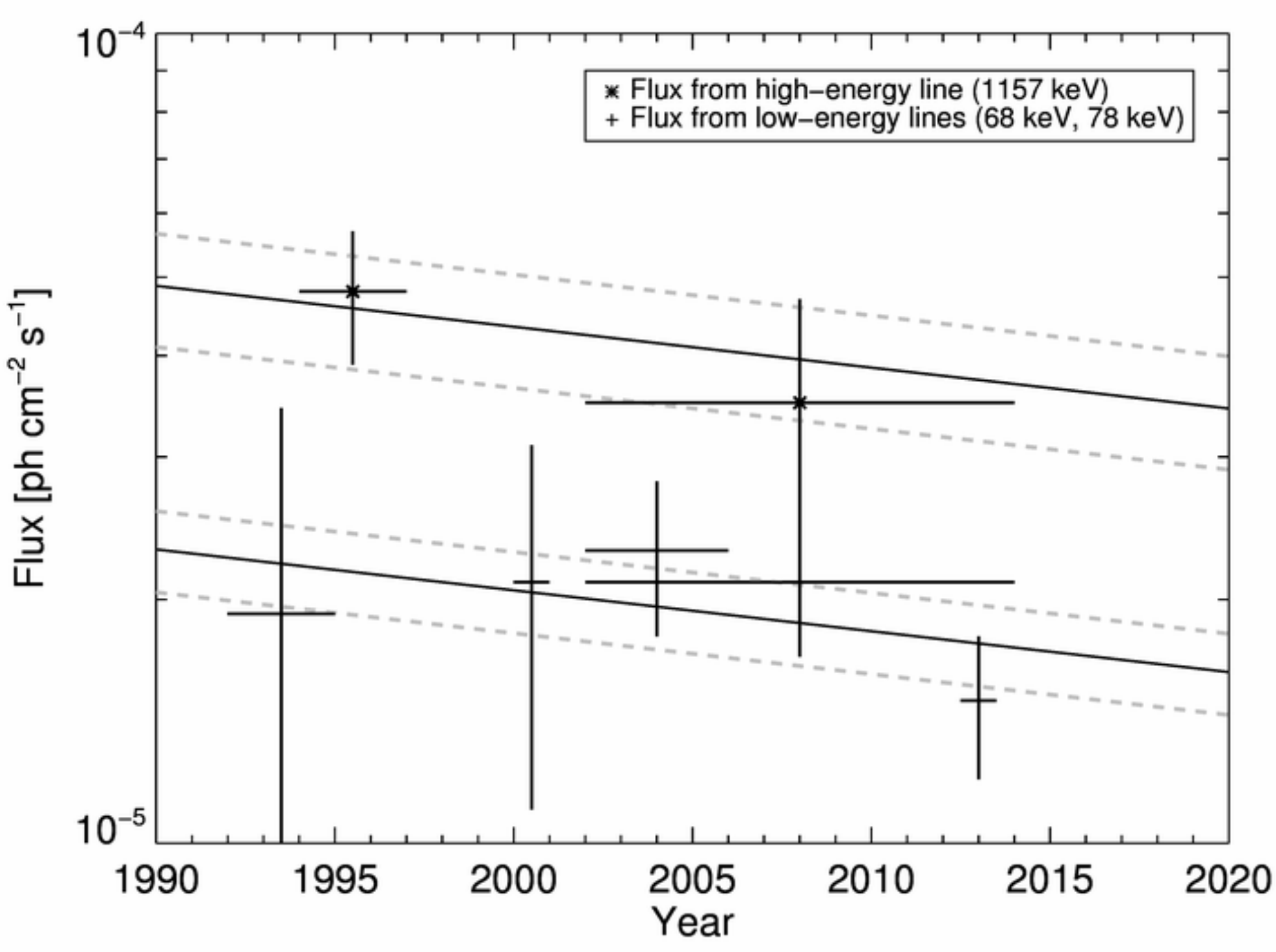} 
\caption{The lines from $^{44}$Ti decay from Cas A, as measured with different instruments (data points with error bars), reflect the decay time of $^{44}$Ti (shown as solid lines). The 1157~keV line from de-excitation of $^{44}$Ca appears somewhat brighter, which could hint towards particle acceleration. (from \cite{Siegert:2015})}
\label{fig_CasA_lines}
\end{figure}%%%%%%%%%%%%%%%%%%%%%%%%%%%%%%%%%%

INTEGRAL's SPI spectrometer recently also added new insight, from re-analysis of archival data \cite{Siegert:2015}. Now the lines from $^{44}$Ti decay at low energy (68 and 78 keV lines originate from the first decay stage and from de-excitation of $^{44}$Sc) as well as the line at 1157~keV originating from de-excitation of $^{44}$Ca have both been measured by the same instrument (Fig.~\ref{fig_CasA_linespectra}). This is important, in view of instrumental systematics, which always is a concern for measurements in the MeV regime and with instruments that suffer from a large instrumental background. Both lines show large Doppler broadening (4300$\pm$1600 / 2200$\pm$1600 km~s$^{-1}$ for the 78 / 1157 keV line, respectively), consistent with the expected ejecta expansion velocity of several 1000~km~s$^{-1}$. The high-energy line appears somewhat brighter than the low-energy line; this brightness ratio also results from comparing all measurements of $^{44}$Ti decay in Cas A since the pioneering COMPTEL detection of the 1157 keV line \cite{Iyudin:1994}, and the various X-ray line measurements thereafter (Fig.~\ref{fig_CasA_lines}) \cite{Vink:2004,Siegert:2015} . Since the brighter line originates from $^{44}$Ca that does not decay any further, it may be possible that nuclear excitation from locally-produced cosmic rays may add to the brightness of the 1157~keV line. Note that young supernova remnants are believed to be the main sources of cosmic rays in the galaxy \cite{Blasi:2013a}; but direct proof has not been possible, as the leakage out of a supernova remnant is expected to occur after typically 1000~years only, when the outer shock weakens, and hence when X- and radio emission from these outer shock regions are difficult to detect. Fermi and H.E.S.S./MAGIC have made measurements of the impacts of cosmic-ray leakage onto the surroundings, in particular molecular clouds, to demonstrate cosmic ray acceleration in supernova remnants; but the difficulty to spectrally distinguish between inverse-Compton acceleration from cosmic ray electrons and pions from cosmic-ray protons has obtained different conclusions for different sources. Low energy cosmic ray interactions such as suggested by this brightness ratio measurement may be searched for, to test this cosmic ray acceleration in a young supernova remnant.

\subsection{Diffuse nucleosynthesis and positrons}
If radioactive decays occur at time scales that are long with respect to the re-occurrence time of nucleosynthesis events, the contributions from individual events will overlap and be super-imposed. Considering that for our Galaxy, the rate of supernovae of type core-collapse is about one supernova every 50 years\cite{Diehl:2006d}, and the rate of novae is about 30-35 events per year, it is clear that all radioactivities beyond the 100 year scale will be of diffuse nature. For the study of such diffuse emission, therefore, it is required that the interpretation accounts for superposition from many sources. The spatial distribution of candidate sources, i.e. their distances and locations within the Galaxy, and also the potential variety in their nucleosynthesis yields have to be modelled from prior knowledge, which can be theory or resulting from other astronomical constraints or combinations thereof. Then, such a source model must be integrated over the radioactive decay time, accounting for temporal evolution of the rate of events, or else, one must adopt a steady state. The latter is plausible and often adopted, i.e. for time scales of My, which relates to the long-lived gamma ray emission sources of $^{26}$Al and $^{60}$Fe and also positron annihilation, one may assume that the event rate in our \emph{current} Galaxy did not change over the past few My, and a steady state has been obtained from the new production and the radioactive decay of unstable nuclei resulting from cosmic nucleosynthesis.

\begin{figure}%[th] %%%%%%%%%%%%%%%%%%%%%%%%%%%%%%%%%%
\centering
\includegraphics[width=0.95\textwidth]{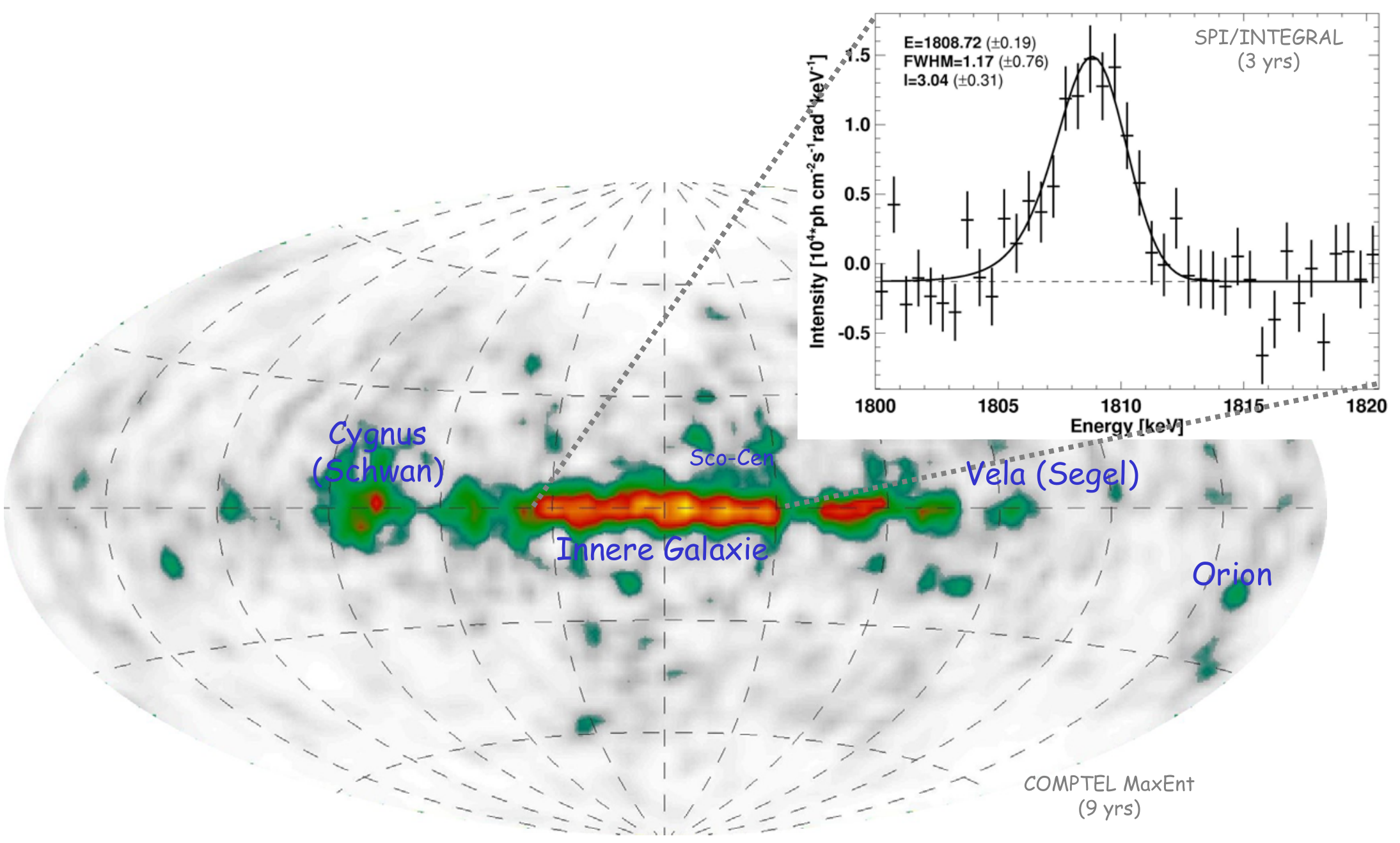} 
\caption{The  $^{26}$Al image derived with COMPTEL \cite{Pluschke:2001c}, and a spectrum from the inner Galaxy as  measured with INTEGRAL/SPI \cite{Diehl:2006c}.  This line reflects current nucleosynthesis in the Galaxy over the past Myrs. 
Its analysis has evolved into a rich field of astrophysics of the structure of our Galaxy, as well as of massive star groups (see text).}
\label{fig_Al-map}
\end{figure}%%%%%%%%%%%%%%%%%%%%%%%%%%%%%%%%%%

\subsubsection{$^{26}$Al}
The discovery of characteristic gamma rays from the radioactive decay to $^{26}$Al   \cite{Mahoney:1982} from the extended inner region of our Galaxy was proof that the synthesis of this isotope, and hence cosmic nucleosynthesis, was ongoing in our Galaxy. This was interpreted as support for above steady-state hypothesis, rather than attributing such $^{26}$Al to a peculiar recent and nearby event (which would have to be located towards the direction of the inner Galaxy, coincidently). Since then, gamma ray surveys have confirmed that $^{26}$Al gamma rays originated from diffuse emission across large regions of the sky, along the plane of our Galaxy. In particular, the COMPTEL data (Fig.~\ref{fig_Al-map}) collected in years 1991-2000 of the CGRO Mission provided support for a sky image that identified the inner Galactic ridge and a few nearby massive star regions as sources of $^{26}$Al. With INTEGRAL's spectrometer, since 2002 data are being taken on the spectral characteristics of this $^{26}$Al emission (Fig.~\ref{fig_Al-map}). This has led to a re-determination of the total content of $^{26}$Al in our Galaxy under the steady-state assumption, and also to the detailed comparison of nucleosynthetic production models considering the source content of localised regions with their population of (mainly massive) stars. Moreover, the line shape measurements also set kinematic constraints to the $^{26}$Al source regions. We will briefly discuss and summarise these aspects in the following.

{\bf The Galactic $^{26}$Al content}.
The determination of the total mass of \Al contained in our Galaxy can be used for a global comparison to models: If one integrates over the different candidate sources, accounting for their yields of interstellar \Al and their spatial distribution (distances), one obtains a general consistency check of our understanding as reflected in models with observational constraints. Early such comparisons \cite{Clayton:1987,Signore:1993} based on data from HEAO-C  \cite{Mahoney:1982} and SMM-GRS \cite{Share:1985} thus saw novae, AGB stars, Wolf-Rayet stars, and core-collapse supernovae as plausible sources, with sufficient uncertainties in yields to regards each of these as significant contributors. The total content of interstellar Galactic \Al was determined as 2--3~\Msol. With COMPTEL's image of the \Al emission across the entire sky \cite{Diehl:1995b,Oberlack:1996,Pluschke:2001c}, it became clear \cite{Prantzos:1996a} that massive stars, i.e. core-collapse supernovae and Wolf-Rayet stars, were dominant contributors, an argument mainly based on the irregular and clumpy appearance of the \Al gamma-ray emission along the plane of the Galaxy, pointing to prominent contributions from clusters of massive stars. These comparisons made use of the bright emission along the inner Galactic ridge, fitting this with intensity-scaled models of spatial source distributions such as disk models based on exponential fall-off with a characteristic scale radius (about 2-4~kpc) and scale height (about 100-300~pc); but also more complex models such as spiral arm representations were used. It became clear, however, that a few more nearby source regions, such as the prominent Cygnus region, or in particular the Scorpius-Centaurus groups at only 110-150~pc distance, could lead to a systematic over-estimate of the Galactic mass of \Al, if not properly accounted for \cite{Martin:2009,Diehl:2010}. 

\begin{figure}%[th] %%%%%%%%%%%%%%%%%%%%%%%%%%%%%%%%%%
\centering
\includegraphics[width=0.6\textwidth]{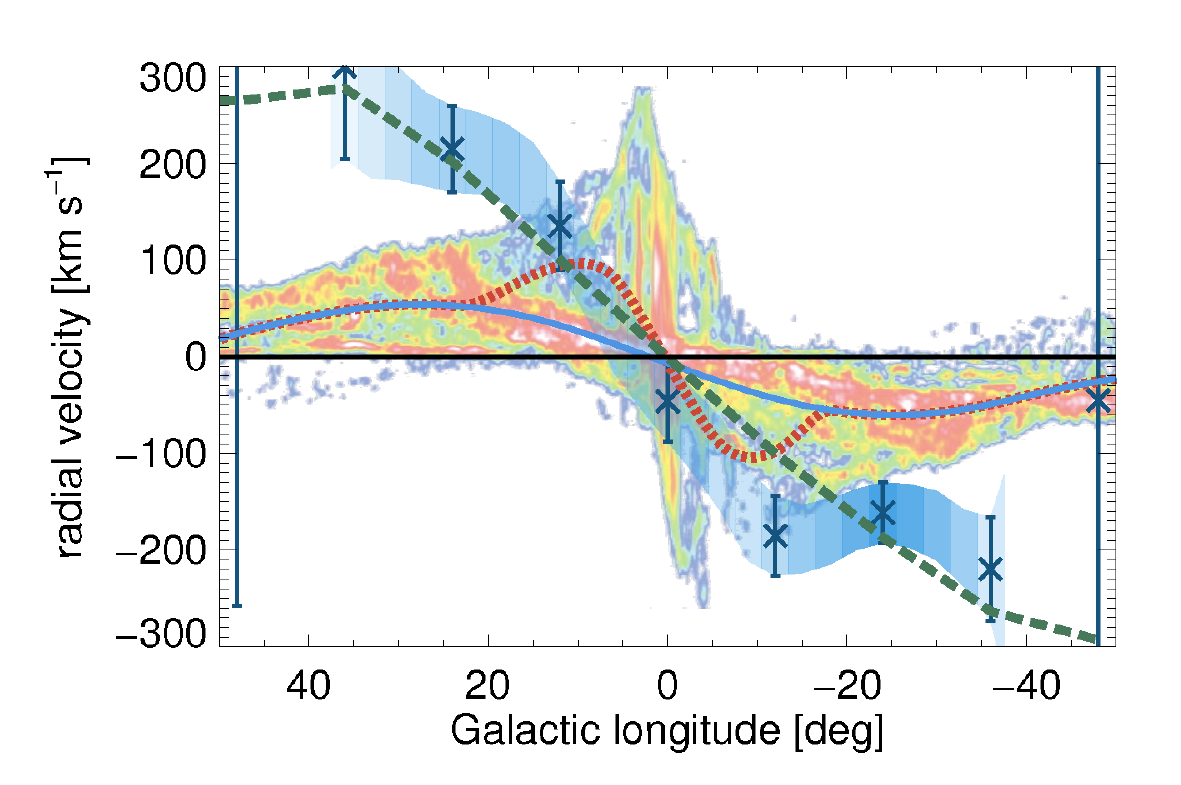} 
\caption{The \Al line shows a clear Doppler shift from large-scale rotation of the Galaxy. The velocities, however, are systematically higher by about 200 km~s$^{-1}$. This is interpreted as \Al being ejected into superbubbles that have expanded more towards the direction forward and away from spiral arms, as the ejection occurs after the groups have evolved over several Myrs and moved away from their formation sites. (From \cite{Kretschmer:2013}).}
\label{fig_26Al_l-v}
\end{figure}%%%%%%%%%%%%%%%%%%%%%%%%%%%%%%%%%%

{\bf $^{26}$Al and massive star groups}.
Predicting the contributions to interstellar \Al from a group of massive stars turned out to be crucial for any bottom-up comparison of source yields with gamma-ray data. Here, a model-predicted yield for a star of mass $M$ must be multiplied with the number of occurrences of stars of this mass, $N(M)$. Then, stellar evolution must be accounted for, as stellar wind phases release \Al much earlier than the core-collapse supernova, which is relevant considering the decay time of \Al of 1.40~My. Population synthesis models were built \cite{Cervino:2002,Voss:2009,Martin:2010b} to put such predictions on a firm footing for comparisons to a range of astronomical messengers, including gamma-ray data, but also interstellar ionisation, interstellar cavity sizes, and stellar emission at other wavelengths. These models mostly used a mass distribution of stars from Salpeter's law, $dN(M)/dM\propto M^{-2.3}$; see however discussion of the present-day mass function aspects  \cite{Weidner:2005,Weidner:2010}. Comparisons were made for specific massive-star groups, where astronomical constraints were available on the stellar census, on interstellar cavities, and on \Al emission. This led to corrections of the massive star content in the Cygnus region \cite{Knodlseder:2000,Martin:2009}, to hints for wind model adjustments from the Carina region \cite{Voss:2012}, and also provided a basis for more detailed investigations for the nearby Orion \cite{Voss:2010a,Diehl:2002,Diehl:2003e,Fierlinger:2012,Fierlinger:2016} and Scorpius-Centaurus \cite{Diehl:2010,Poppel:2010,Preibisch:2002} groups, where astronomical constraints are considered rather complete. These studies therefore are ongoing, and allow, a.o., for testing stellar evolution models with and without inclusion of stellar rotation, for different mass functions including non-explosion of some mass intervals in the 10--30~\Msol range, stellar wind and supernova overlaps in superbubbles, and correlated star formation scenarios (triggered star formation).

{\bf Implications of the Galactic $^{26}$Al line characteristics}.
INTEGRAL's spectrometer has been maintaining the excellent spectral resolution of its Ge detectors over more than 13 mission years now, through periodic annealing operations of the detectors. Here, roughly every 6 months, the cryo-cooled Ge detector array with its 19 elements is elevated from normal operation temperatures ($\sim$80~K) to $\sim$10$^o$~C for about two weeks, thus curing the lattice defects in Ge detectors which compromise charge collection, and are created by cosmic-ray bombardments in space. As a result, the survey data of INTEGRAL over many years can be analysed coherently at high spectral resolution. This allowed to recognise the large-scale Galactic rotation in the \Al line through its characteristic systematic Doppler shifts with Galactic longitude viewing direction (Fig.~\ref{fig_26Al_l-v}) \cite{Diehl:2006d,Kretschmer:2013}.  These measurements led to a precision of line shape constraints, which allowed to compare the large-scale Galactic rotation velocities with longitude between \Al, which traces nucleosynthesis ejecta, and CO, which traces molecular clouds parental to new star formation and massive star groups \cite{Kretschmer:2013}. The apparent discrepancy of velocities, with an excess velocity of $\sim$200~km~s$^{-1}$ of \Al-based values over CO-based values, came as a surprise. The explanation was obtained from modelling how massive star groups form, evolve, and dissolve over tens of My, recognising that the \Al ejection mostly will occur into cavities created from stellar winds and earlier star formation, and that such cavities are more likely to occur between spiral arms than within them. Therefore, superbubbles are recognised \cite{Krause:2015} through these \Al measurements as major elements in the process of returning nucleosynthesis ejecta from stars and supernovae into cold interstellar gas that may later form next-generation stars. We have tested several aspects of this \emph{stellar feedback} with respect to X-ray emission \cite{Krause:2014a} and the origins of globular clusters with their peculiar abundance patterns \cite{Krause:2016}.  At the galactic scale, superbubbles will also influence how the halo and interstellar clouds herein are involved into the cycling of matter from star forming region to next-generation star forming region \cite{Fraternali:2012}, and how disk and halo exchange angular momentum \cite{Kretschmer:2013,Krause:2015}.

\begin{figure}%[th] %%%%%%%%%%%%%%%%%%%%%%%%%%%%%%%%%%
\centering
\includegraphics[width=0.6\textwidth]{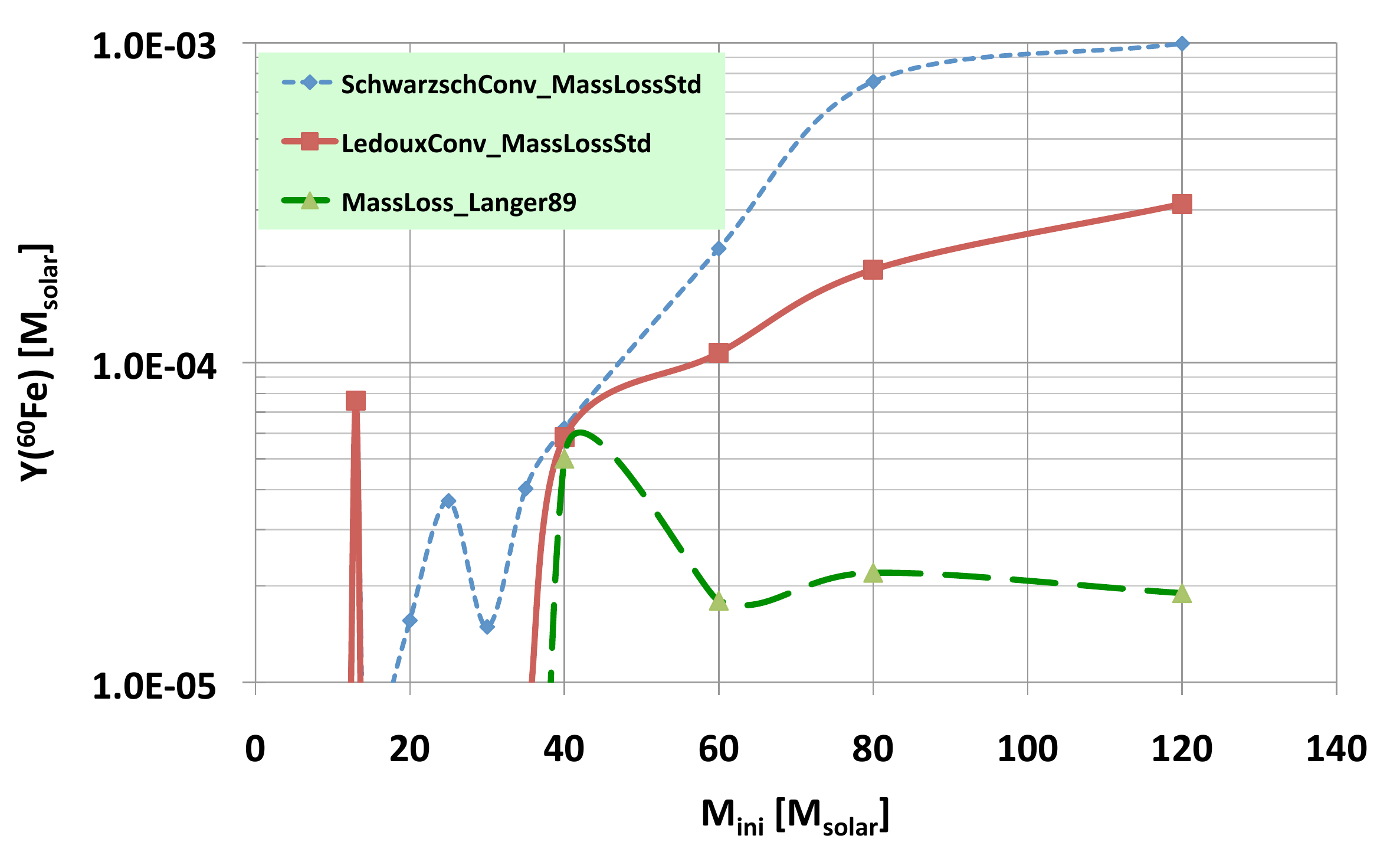} 
\caption{The yields in $^{60}$Fe from massive stars are rather uncertain. In this graph, adapted from \cite{Limongi:2006qf}, we illustrate the impact of different treatments of known physical processes inside the star. }
\label{fig_60Feyields}
\end{figure}%%%%%%%%%%%%%%%%%%%%%%%%%%%%%%%%%%

\subsubsection{$^{60}$Fe}
The gamma-ray emission from $^{60}$Fe was predicted from nucleosynthesis theory \cite{Clayton:1971}. This was re-affirmed after $^{26}$Al has been found, and the massive star source hypothesis turned out to be more plausible and dominant as compared to a nova origin \cite{Timmes:1995}. Massive star interiors are candidate sites of neutron capture on iron nuclei and thus $^{60}$Fe producers, in particular in the He and C shells where neutron release from the $^{13}$C($\alpha$,n) and $^{22}$Ne($\alpha$,n) reactions were expected \cite{Limongi:2006}. It just remained unclear how much of the $^{60}$Fe would survive and not be destroyed by further neutron captures. Predictions range from insignificant amounts to amounts that rival $^{26}$Al yields, yet strongly depend on uncertainties of massive star shell burning astrophysics as well as the neutron capture and beta decay reaction rates involved (see Fig. \ref{fig_60Feyields}). The shell burning stages within massive stars in their late evolution are quite uncertain: Here, 3D effects of gas motions from shellular rotation stimulated by the rotating star play a key role, as well

Around 2005, gamma rays from $^{60}$Fe decays in the Galaxy were reported from both the RHESSI \cite{Harris:2005} and the INTEGRAL \cite{Harris:2004} observations of our Galaxy. The early reports were not highly significant detections, but after more INTEGRAL observations seemed consolidated \cite{Wang:2007a}. It became clear that the gamma ray brightness of $^{60}$Fe was far below the brightness in $^{26}$Al, the intensity ratio being about 15\% ($\pm$5\%). Note that $\beta$-decay lifetime measurements as well as neutron capture measurements are far from simple, and recent efforts have contributed here \cite{Rugel:2009,Heftrich:2015}. This allows some interesting astrophysical constraints, yet precludes a detailed study of line shapes and regional/localised emission regions, which had turned $^{26}$Al studies into such a rich field in the meantime.

\begin{figure}%[th] %%%%%%%%%%%%%%%%%%%%%%%%%%%%%%%%%%
\centering
\includegraphics[width=0.33\textwidth]{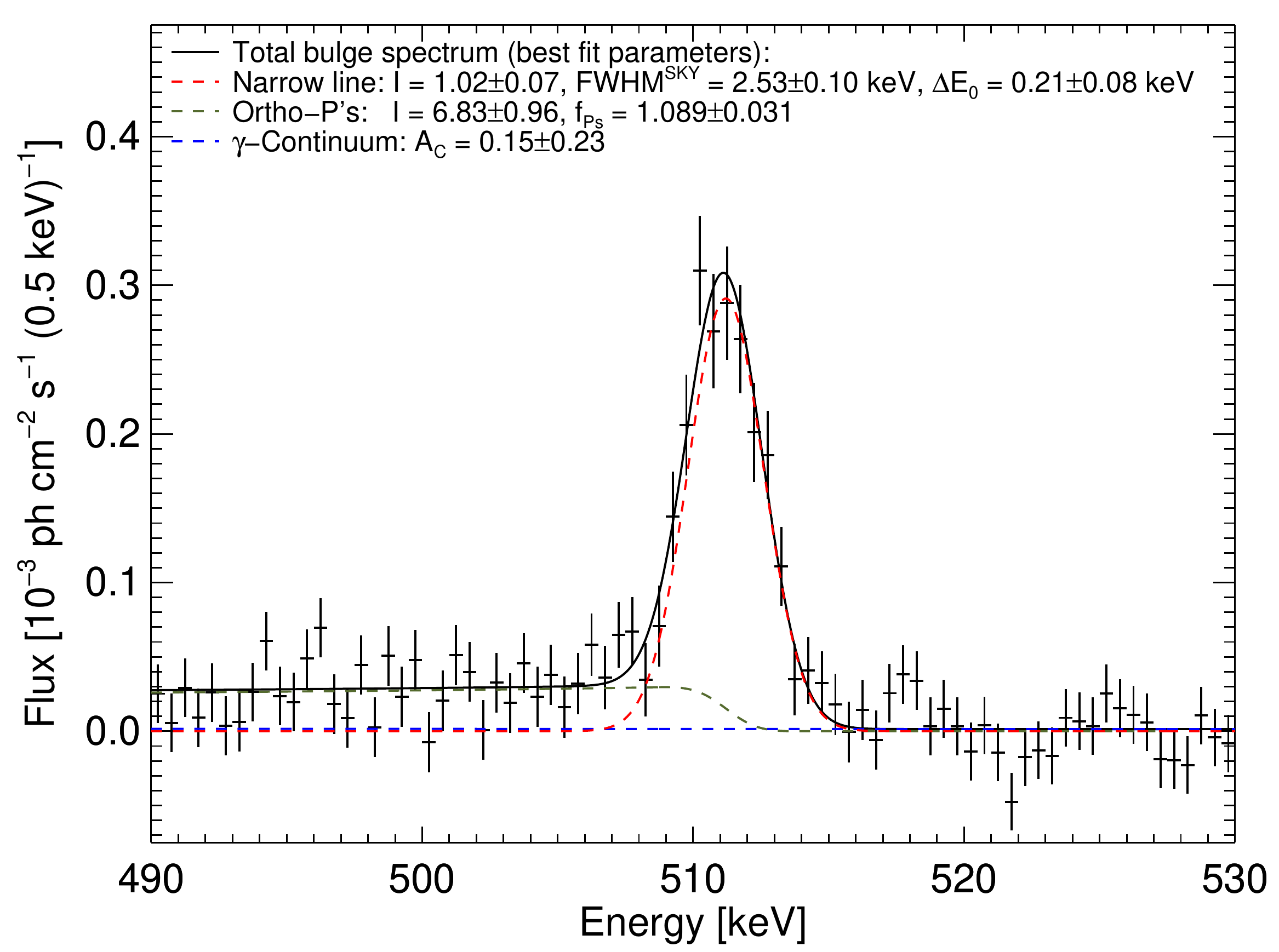} %\break
\includegraphics[width=0.33\textwidth]{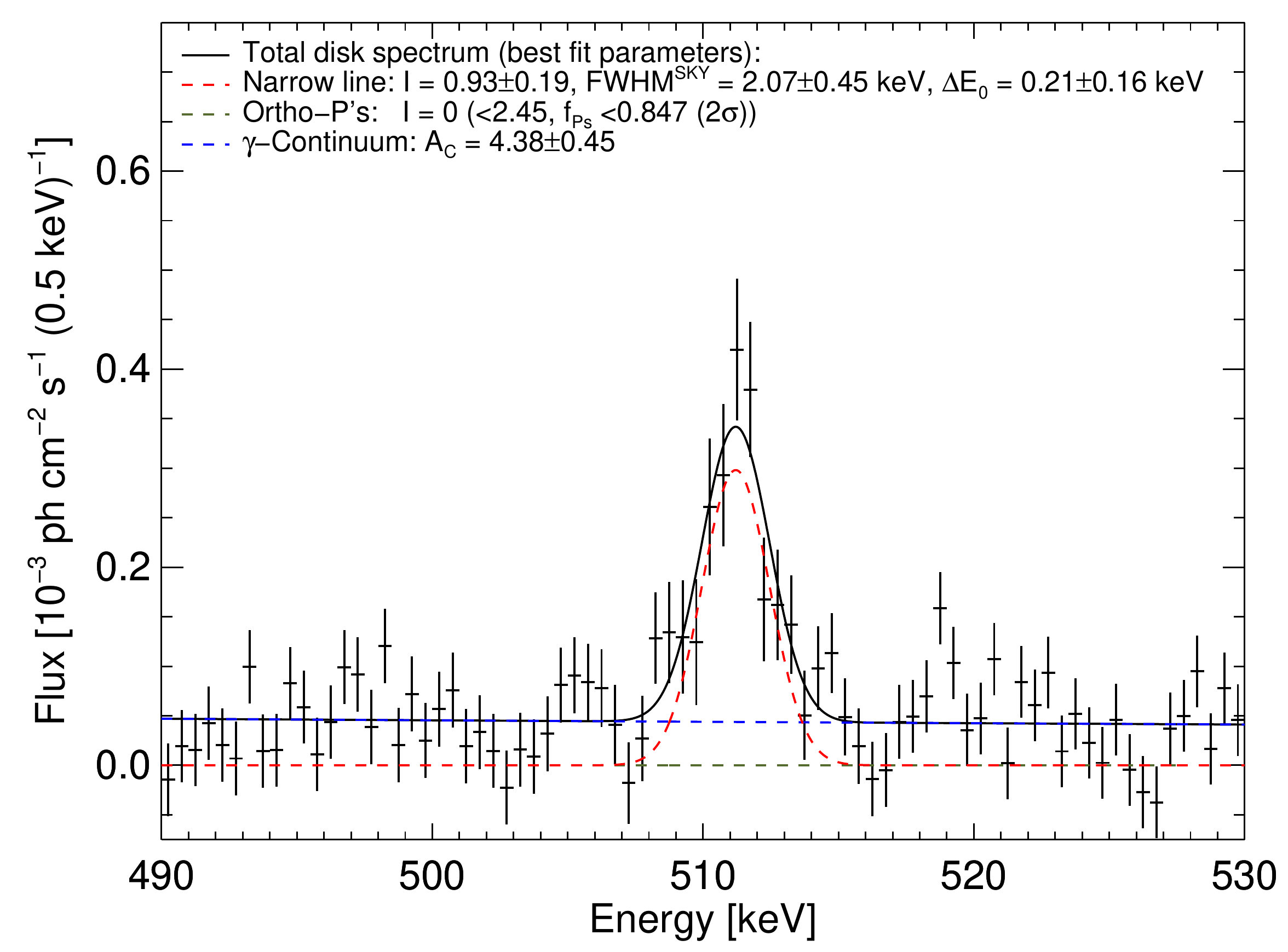} %\break
\includegraphics[width=0.33\textwidth]{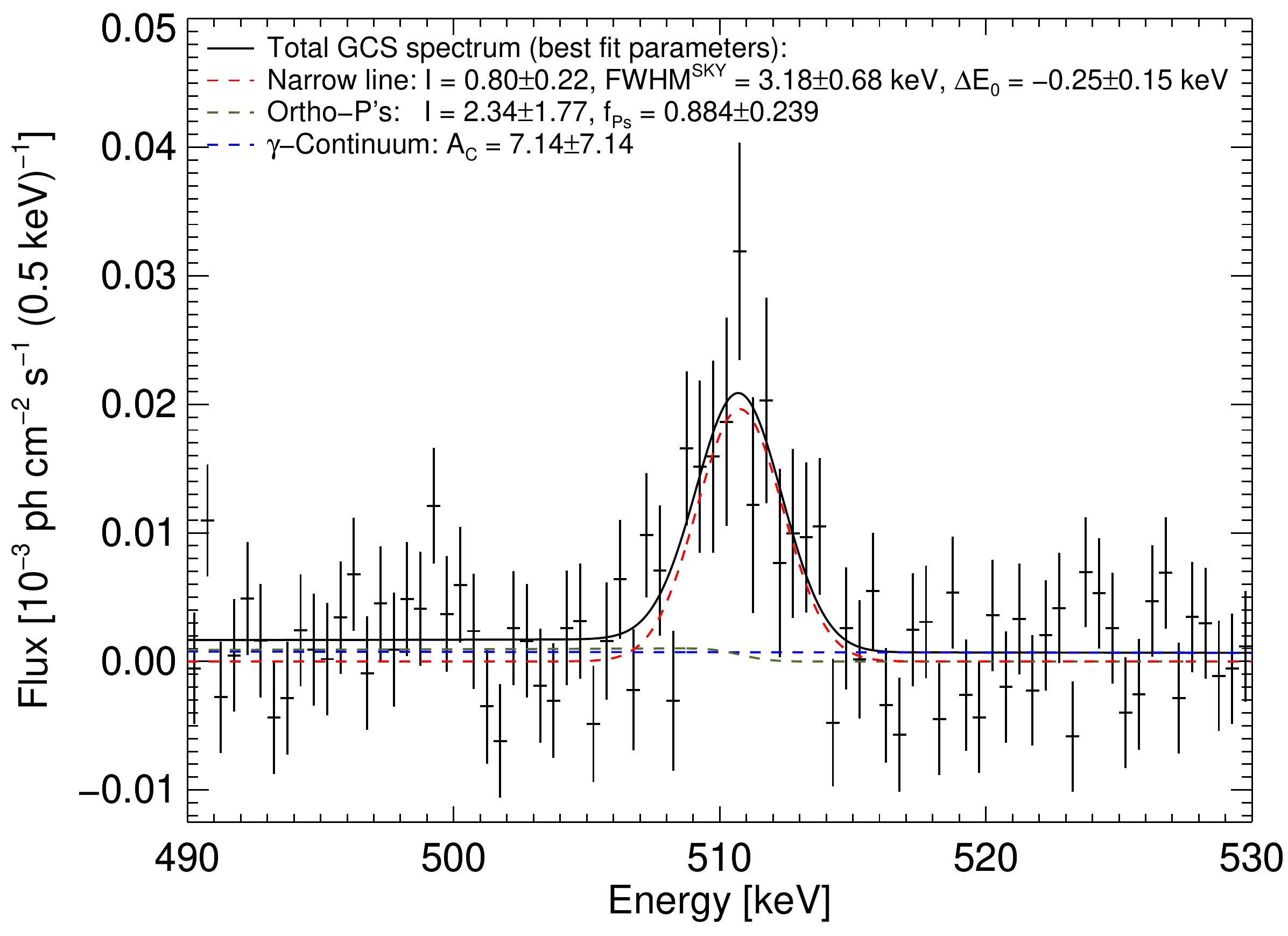} 
\caption{The spectra obtained from separating contributions from the bright bulge (\emph{top left}), the extended disk (\emph{top right}), and a potential point source in the central Galaxy (\emph{bottom}) (from \cite{Siegert:2016}). }
\label{fig_511Spectra}
\end{figure}%%%%%%%%%%%%%%%%%%%%%%%%%%%%%%%%%%

\subsubsection{Positron annihilation}
Positron emission results from the production of proton rich isotopes, in general. This occurs mainly in hydrogen burning in stars and nova explosions. Thus positron annihilation may be plausibly expected to occur after nova explosions, but also in type-I X-ray bursting sources in neutron star binaries; in normal stars, nucleosynthesis release times into interstellar space are longer than any of the decay times of $\beta^+$ radioactivities. Therefore, the study of positron annihilation across our Galaxy also is related to the field of nuclear astrophysics, and the contribution to positrons from nucleosynthesis have been studied since early times \cite{Clayton:1973,Lingenfelter:1978,Ramaty:1994}. A recent assessment \cite{Martin:2012}, including the propagation of positrons from their nucleosynthesis sources to potential annihilation sites, has updated these expected contributions to positron annihilation gamma rays from nucleosynthesis sources.

Beyond the connection to nucleosynthesis and $\beta^+$-unstable isotopes, positrons may also be produced from other objects, such as pair plasma creating high-energy sources (pulsars, accreting binaries), and by cosmic ray interactions in interstellar space \cite{Prantzos:2011}. 

Already from OSSE results from the Compton Gamma Ray Observatory mission, it had been suggested that the annihilation gamma-ray emission was diffuse in nature, but particularly bright in the region of the inner Galaxy and its bulge region \cite{Purcell:1997}. These data had ruled out the transient annihilation hypothesis based on the 'Great Annihilator' results discussed earlier \cite{Leventhal:1989,Ramaty:1992,Mirabel:1992}.  
INTEGRAL observations then have provided a rather detailed picture from the gamma ray sky in positron annihilation emission\cite{Jean:2003a,Knodlseder:2005,Churazov:2005,Jean:2006,Weidenspointner:2008a,Bouchet:2010,Skinner:2012}. It now was clear that the inner Galaxy with its bulge region was a very prominent source of annihilation gamma rays, much brighter than the disk region throughout the Galaxy itself. This was surprising, as practically all candidate sources would be expected to populate the Galactic disk . Also surprising was the rather regular appearance of the central Galaxy emission, slightly offset from longitude zero by about 1 degree, but otherwise rather smooth in distribution and circularly extended both in longitude and latitude, suggesting an extent of 1--2~kpc. This stimulated interpretations of dark matter as a main source, as in this case the Galaxy's gravitational potential would dictate the spatial distribution of the emission. However, the \Al emission along the plane of the Galaxy demonstrates that positron sources in the Galactic disk regions do exist, and should contribute to the gamma-ray emission.

After more than four years of data collection, INTEGRAL began to reveal the disk emission in positron annihilation gamma-rays as well \cite{Weidenspointner:2008a,Bouchet:2010}. Some discussion about asymmetries ensued \cite{Weidenspointner:2008a}, and was later understood as originating from the above-mentioned slight offset of the emission centroid \cite{Bouchet:2010}, while analyses searched for differences in brightness at positive versus negative longitudes throughout the Galaxy \cite{Skinner:2012}. 
Recently, from 11 years of data, the spatial distribution and spectroscopy has been re-assessed (Fig.~\ref{fig_511Spectra}) \cite{Siegert:2016}. In those latest results it became clear that the disk also is a substantial emitter of positron annihilation gamma rays, rivalling the bulge region in total brightness, but with much lower surface brightness. Therefore it had been difficult to detect this disk emission with less exposure in earlier data. Also, the disk appears now to be more extended in latitude than believed earlier, reducing the surface brightness even more. This suggests that positrons do escape from source regions also towards the halo, thus leading to increased latitude extent of the gamma ray emission. Possibly, positrons escape into the halo, and may then be re-directed into the inner Galaxy and bulge region by large-scale magnetic fields in the halo, following an earlier model to explain the bright bulge emission.

The annihilation of positrons may occur after slowing down from the relativistic energies at which they typically are produced, and only at energies in the eV region. Then, in interstellar space the formation of positronium through charge-exchange collisions with hydrogen is important, and an intermediate stage that eases annihilation because of angular momentum conservation. Positronium is an atom-like bound quantum mechanical system composed of the positron and an electron, with a binding energy of 6.8 eV. From its relative spin orientations, it can adopt a singlet and a triplet status. Annihilation then can occur from either of these states, the triplet state being somewhat more long-lived and producing three photons from angular momentum conservations, while the singlet state produces two photons of 511 keV energy. Annihilation thus converts the electron-positron rest masses of 1.022~MeV into radiation, producing a line at 511 keV and a characteristic continuum towards lower energies. The relative intensities of these two spectral components encode how much annihilation occurs through the positronium intermediate state, versus direct annihilation that is the only annihilation channel in fully-ionised interstellar plasma. From this, it had been deduced that positrons typically annihilate in a partially-ionised medium with sufficient neutral atoms to make positronium annihilation the dominant channel, and from the width of the 511~keV line the temperature of the medium is constrained through Doppler broadening to about 7000~K \cite{Churazov:2005,Churazov:2015}.   In latest results  \cite{Siegert:2016}, the different spectral components and line width could be identified separately in emission attributed to the bright bulge region, to the extended and faint disk region, and to another point-like source near the Galactic Centre direction. It appears that spectra may differ between bulge and disk  (Fig.~\ref{fig_511Spectra}, upper spectra) , indicating different typical/averaged annihilation conditions. Also, there are indications that the central source shows a broader 511~keV line  (Fig.~\ref{fig_511Spectra}, below) , which may suggest that annihilations here occur in gas with higher kinematic motion or at higher temperature, such as expected if the central black hole environment was related to this source.

\begin{figure}%[th] %%%%%%%%%%%%%%%%%%%%%%%%%%%%%%%%%%
\centering
\includegraphics[width=0.7\textwidth]{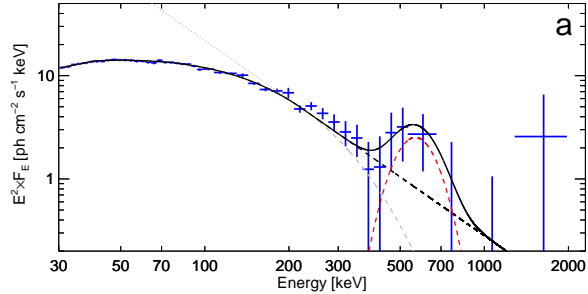} 
\caption{The spectrum of microquasar V404 Cygni in its early flaring period June 2015, obtained from INTEGRAL/SPI observations (from \cite{Siegert:2016a}). Here, the annihilation occurs in hot plasma clouds near the black hole, kinematic shift and broadening from the Doppler effect results in the bumb feature between 0.5 and 1 MeV, consistent with theoretical expectations. }
\label{fig_511SpectraV404}
\end{figure}%%%%%%%%%%%%%%%%%%%%%%%%%%%%%%%%%%

One of the candidate positron sources are accreting binary systems with accretion onto a stellar-mass black hole; these objects are called micro-quasars  \cite{Prantzos:2011}. Here, the matter is channeled towards the black hole through an  accretion disk that shines in X-rays \cite{Done:2007,Remillard:2006}, and disappears in the black hole, somehow resulting in plasma ejections seen as jets of radio emission far out in those systems \cite{Mirabel:1999}. Such binary system accretion often is unstable, reflecting different configurations of the accretion flow and the geometrical shapes of the accretion disk. 
One of these microquasars, V404 Cygni, nearby in our Galaxy, went into a flaring state in June 2015 \cite{Rodriguez:2015}, thus becoming the brightest source in the X-ray sky for several weeks. Analysing INTEGRAL data from this flaring period, it was recently recognised \cite{Siegert:2016a} that the typical gamma-ray signatures of positron annihilation were present  (Fig.~\ref{fig_511SpectraV404}), although significantly Doppler broadened and variable, and thus not easy to detect, typically. This underlines the candidate source role of microquasars to produce positrons which later annihilate in interstellar space. The local annihilation seen recently with V404 Cygni occurs near the production site very close to the black hole itself, and could be seen in this source only because it is bright enough to also have sufficient brightness in such a side channel of its total radiation spectrum.

\section{Summary and perspectives}
The measurement of cosmic gamma rays in the nuclear domain, from $\sim$100~keV to a few MeV, have evolved into a diverse and rather fruitful field of high energy astronomy. Its science is rather unique, addressing cosmic sources of nucleosynthesis, of particle acceleration, and of positron production. Particle acceleration can also be studied at higher energies in gamma rays, and with cosmic rays. The other fields are unique to this spectral window. 
Current instrumentation features Compton scattering or coded mask techniques for spatial resolution, and semiconductors for spectral precision. Through better segmentation of detectors and tracking of secondary particles, the precision and also sensitivity could be advanced now to explore sources beyond our Galaxy and many expected but fainter Galactic sources. But space missions are costly and sparse.
INTEGRAL's spectrometer, together with the NuSTAR hard X-ray telescope, are the only current missions which provide measurements in this energy range. No further such instruments are currently planned, and therefore will probably not be realised  within the next decades (space projects typically have 10-20 years of lead time). But INTEGRAL is working well after 13 years, even though it was originally planned for a 5-year maximum of operations. Its orbit and onboard resources should allow proper functionality until 2029, when the spacecraft will re-enter the Earth atmosphere and be de-orbited, unless some unforeseen malfunction or micrometeorite collision would occur earlier. Re-approval after science and operations review occurs every two years at ESA, and currently the mission is financed till end 2018. As above results show, important scientific refinements of gamma-ray results occur in the mature and late INTEGRAL mission, and transient sources such as SN2014J and V404 Cygni make these mission extensions more worthwhile.  

\section*{Acknowledgments}
The INTEGRAL/SPI project has been completed under the responsibility and leadership of CNES; we are grateful to ASI, CEA, CNES, DLR, ESA, INTA, NASA and OSTC for support of
this ESA space science mission. R.D. acknowledges support from the Munich cluster of excellence \emph{origin and evolution of the universe}, and from the organisers of this summer school.

%%%%%%%%%%%%%%%%%%%%%%%%%%%%%%%%%%%%%%%%%%%
%\section{References}

%%%%%%%%%%%%%%%%%%%%%%%%%%%%%%%%% satisfy BibTeX from ADS %%%%%%%%%%%%%%

          % Astronomical Journal
\newcommand{\actaa}{Acta Astron.}%
  % Acta Astronomica
\newcommand{\araa}{ARA\&A}%
          % Annual Review of Astron and Astrophys
\newcommand{\apj}{ApJ}%
          % Astrophysical Journal
\newcommand{\apjl}{ApJ}%
          % Astrophysical Journal, Letters
\newcommand{\apjs}{ApJS}%
          % Astrophysical Journal, Supplement
\newcommand{\ao}{Appl.~Opt.}%
          % Applied Optics
\newcommand{\apss}{Ap\&SS}%
          % Astrophysics and Space Science
\newcommand{\aap}{Astron.\&Astroph.}%
          % Astronomy and Astrophysics
\newcommand{\aapr}{Astron.\&Astroph.~Rev.}%
          % Astronomy and Astrophysics Reviews
\newcommand{\aaps}{Astron.\&Astroph.~Suppl.}%
          % Astronomy and Astrophysics, Supplement
\newcommand{\aj}{Astron.Journ.}%
          % Astronomical Journal
\newcommand{\azh}{AZh}%
          % Astronomicheskii Zhurnal
\newcommand{\memras}{MmRAS}%
          % Memoirs of the RAS
\newcommand{\mnras}{Mon.Not.Royal~Astr.~Soc.}%
          % Monthly Notices of the RAS
\newcommand{\na}{New Astron.}%
  % New Astronomy
\newcommand{\nar}{New Astron. Rev.}%
  % New Astronomy Review
\newcommand{\pra}{Phys.~Rev.~A}%
          % Physical Review A: General Physics
\newcommand{\prb}{Phys.~Rev.~B}%
          % Physical Review B: Solid State
\newcommand{\prc}{Phys.~Rev.~C}%
          % Physical Review C
\newcommand{\prd}{Phys.~Rev.~D}%
          % Physical Review D
\newcommand{\pre}{Phys.~Rev.~E}%
          % Physical Review E
\newcommand{\prl}{Phys.~Rev.~Lett.}%
          % Physical Review Letters
\newcommand{\pasa}{PASA}%
  % Publications of the Astron. Soc. of Australia
\newcommand{\pasp}{PASP}%
          % Publications of the ASP
\newcommand{\pasj}{PASJ}%
          % Publications of the ASJ
\newcommand{\skytel}{S\&T}%
          % Sky and Telescope
\newcommand{\solphys}{Sol.~Phys.}%
          % Solar Physics
\newcommand{\sovast}{Soviet~Ast.}%
          % Soviet Astronomy
\newcommand{\ssr}{Space~Sci.~Rev.}%
          % Space Science Reviews
\newcommand{\nat}{Nature}%
          % Nature
\newcommand{\iaucirc}{IAU~Circ.}%
          % IAU Cirulars
\newcommand{\aplett}{Astrophys.~Lett.}%
          % Astrophysics Letters and Communications
\newcommand{\apspr}{Astrophys.~Space~Phys.~Res.}%
          % Astrophysics Space Physics Research
\newcommand{\nphysa}{Nucl.~Phys.~A}%
          % Nuclear Physics A
\newcommand{\physrep}{Phys.~Rep.}%
          % Physics Reports
\newcommand{\procspie}{Proc.~SPIE}%
          % Proceedings of the SPIE

 % define Journal name commands as in bib

\bibliographystyle{plain}

%\begin{thebibliography}{999}
%%\bibliography{rod-refs16,rod-refs_SNIa}

\begin{thebibliography}{10}

\bibitem{Bertsch:1988}
D.~L. {Bertsch}, C.~E. {Fichtel}, and J.~I. {Trombka}.
\newblock {Instrumentation for gamma-ray astronomy}.
\newblock {\em \ssr}, 48:113--168, March 1988.

\bibitem{Blasi:2013a}
P.~{Blasi}.
\newblock {The origin of galactic cosmic rays}.
\newblock {\em \aapr}, 21:70, November 2013.

\bibitem{Boggs:2015}
S.~E. {Boggs}, F.~A. {Harrison}, H.~{Miyasaka}, B.~W. {Grefenstette},
  A.~{Zoglauer}, C.~L. {Fryer}, S.~P. {Reynolds}, D.~M. {Alexander}, H.~{An},
  D.~{Barret}, F.~E. {Christensen}, W.~W. {Craig}, K.~{Forster}, P.~{Giommi},
  C.~J. {Hailey}, A.~{Hornstrup}, T.~{Kitaguchi}, J.~E. {Koglin}, K.~K.
  {Madsen}, P.~H. {Mao}, K.~{Mori}, M.~{Perri}, M.~J. {Pivovaroff},
  S.~{Puccetti}, V.~{Rana}, D.~{Stern}, N.~J. {Westergaard}, and W.~W. {Zhang}.
\newblock {$^{44}$Ti gamma-ray emission lines from SN1987A reveal an asymmetric
  explosion}.
\newblock {\em Science}, 348:670--671, May 2015.

\bibitem{BOREXINO-Collaboration:2008}
{BOREXINO Collaboration}, C.~{Arpesella}, G.~{Bellini}, J.~{Benziger},
  S.~{Bonetti}, B.~{Caccianiga}, F.~{Calaprice}, F.~{Dalnoki-Veress},
  D.~{D'Angelo}, H.~{de Kerret}, A.~{Derbin}, M.~{Deutsch}, A.~{Etenko},
  K.~{Fomenko}, R.~{Ford}, D.~{Franco}, B.~{Freudiger}, C.~{Galbiati},
  S.~{Gazzana}, M.~{Giammarchi}, M.~{Goeger-Neff}, A.~{Goretti}, C.~{Grieb},
  S.~{Hardy}, G.~{Heusser}, A.~{Ianni}, A.~{Ianni}, M.~{Joyce}, G.~{Korga},
  D.~{Kryn}, M.~{Laubenstein}, M.~{Leung}, E.~{Litvinovich}, P.~{Lombardi},
  L.~{Ludhova}, I.~{Machulin}, G.~{Manuzio}, A.~{Martemianov}, F.~{Masetti},
  K.~{McCarty}, E.~{Meroni}, L.~{Miramonti}, M.~{Misiaszek}, D.~{Montanari},
  M.~E. {Monzani}, V.~{Muratova}, L.~{Niedermeier}, L.~{Oberauer},
  M.~{Obolensky}, F.~{Ortica}, M.~{Pallavicini}, L.~{Papp}, L.~{Perasso},
  A.~{Pocar}, R.~S. {Raghavan}, G.~{Ranucci}, A.~{Razeto}, A.~{Sabelnikov},
  C.~{Salvo}, S.~{Sch{\"o}nert}, H.~{Simgen}, O.~{Smirnov}, M.~{Skorokhvatov},
  A.~{Sonnenschein}, A.~{Sotnikov}, S.~{Sukhotin}, Y.~{Suvorov},
  V.~{Tarasenkov}, R.~{Tartaglia}, G.~{Testera}, D.~{Vignaud}, S.~{Vitale},
  R.~B. {Vogelaar}, F.~{von Feilitzsch}, M.~{Wojcik}, O.~{Zaimidoroga},
  S.~{Zavatarelli}, and G.~{Zuzel}.
\newblock {First real time detection of $^{7}$Be solar neutrinos by Borexino}.
\newblock {\em Physics Letters B}, 658:101--108, January 2008.

\bibitem{BOREXINO-Collaboration:2014}
{BOREXINO Collaboration}, G.~{Bellini}, J.~{Benziger}, D.~{Bick}, G.~{Bonfini},
  D.~{Bravo}, B.~{Caccianiga}, L.~{Cadonati}, F.~{Calaprice}, A.~{Caminata},
  P.~{Cavalcante}, A.~{Chavarria}, A.~{Chepurnov}, D.~{D'Angelo}, S.~{Davini},
  A.~{Derbin}, A.~{Empl}, A.~{Etenko}, K.~{Fomenko}, D.~{Franco},
  F.~{Gabriele}, C.~{Galbiati}, S.~{Gazzana}, C.~{Ghiano}, M.~{Giammarchi},
  M.~{G{\"o}ger-Neff}, A.~{Goretti}, M.~{Gromov}, C.~{Hagner}, E.~{Hungerford},
  A.~{Ianni}, A.~{Ianni}, V.~{Kobychev}, D.~{Korablev}, G.~{Korga}, D.~{Kryn},
  M.~{Laubenstein}, B.~{Lehnert}, T.~{Lewke}, E.~{Litvinovich}, F.~{Lombardi},
  P.~{Lombardi}, L.~{Ludhova}, G.~{Lukyanchenko}, I.~{Machulin}, S.~{Manecki},
  W.~{Maneschg}, S.~{Marcocci}, Q.~{Meindl}, E.~{Meroni}, M.~{Meyer},
  L.~{Miramonti}, M.~{Misiaszek}, M.~{Montuschi}, P.~{Mosteiro}, V.~{Muratova},
  L.~{Oberauer}, M.~{Obolensky}, F.~{Ortica}, K.~{Otis}, M.~{Pallavicini},
  L.~{Papp}, L.~{Perasso}, A.~{Pocar}, G.~{Ranucci}, A.~{Razeto}, A.~{Re},
  A.~{Romani}, N.~{Rossi}, R.~{Saldanha}, C.~{Salvo}, S.~{Sch{\"o}nert},
  H.~{Simgen}, M.~{Skorokhvatov}, O.~{Smirnov}, A.~{Sotnikov}, S.~{Sukhotin},
  Y.~{Suvorov}, R.~{Tartaglia}, G.~{Testera}, D.~{Vignaud}, R.~B. {Vogelaar},
  F.~{von Feilitzsch}, H.~{Wang}, J.~{Winter}, M.~{Wojcik}, A.~{Wright},
  M.~{Wurm}, O.~{Zaimidoroga}, S.~{Zavatarelli}, K.~{Zuber}, and G.~{Zuzel}.
\newblock {Neutrinos from the primary proton-proton fusion process in the Sun}.
\newblock {\em \nat}, 512:383--386, August 2014.

\bibitem{Bouchet:2010}
L.~{Bouchet}, J.~P. {Roques}, and E.~{Jourdain}.
\newblock {On the Morphology of the Electron-Positron Annihilation Emission as
  Seen by SPI/INTEGRAL}.
\newblock {\em \apj}, 720:1772--1780, September 2010.

\bibitem{Cervino:2002}
M.~{Cervi{\~n}o}, R.~{Diehl}, K.~{Kretschmer}, and S.~{Pl{\"u}schke}.
\newblock {Radioactive isotopes in star forming regions}.
\newblock {\em New Astronomy Review}, 46:541--545, July 2002.

\bibitem{Churazov:2015}
E.~{Churazov}, R.~{Sunyaev}, J.~{Isern}, I.~{Bikmaev}, E.~{Bravo}, N.~{Chugai},
  S.~{Grebenev}, P.~{Jean}, J.~{Kn{\"o}dlseder}, F.~{Lebrun}, and
  E.~{Kuulkers}.
\newblock {Gamma-rays from Type Ia Supernova SN2014J}.
\newblock {\em \apj}, 812:62, October 2015.

\bibitem{Churazov:2014a}
E.~{Churazov}, R.~{Sunyaev}, J.~{Isern}, J.~{Kn{\"o}dlseder}, P.~{Jean},
  F.~{Lebrun}, N.~{Chugai}, S.~{Grebenev}, E.~{Bravo}, S.~{Sazonov}, and
  M.~{Renaud}.
\newblock {Cobalt-56 {$\gamma$}-ray emission lines from the type Ia supernova
  2014J}.
\newblock {\em \nat}, 512:406--408, August 2014.

\bibitem{Churazov:2005}
E.~{Churazov}, R.~{Sunyaev}, S.~{Sazonov}, M.~{Revnivtsev}, and
  D.~{Varshalovich}.
\newblock {Positron annihilation spectrum from the Galactic Centre region
  observed by SPI/INTEGRAL}.
\newblock {\em \mnras}, 357:1377--1386, March 2005.

\bibitem{Clayton:1971}
D.~D. {Clayton}.
\newblock {New Prospect for Gamma-Ray-Line Astronomy}.
\newblock {\em \nat}, 234:291--+, December 1971.

\bibitem{Clayton:1973}
D.~D. {Clayton}.
\newblock {Galaxy-Positronium origin of 476 keV galactic feature}.
\newblock {\em \nat}, 244:137--+, August 1973.

\bibitem{Clayton:1969}
D.~D. {Clayton}, S.~A. {Colgate}, and G.~J. {Fishman}.
\newblock {Gamma-Ray Lines from Young Supernova Remnants}.
\newblock {\em \apj}, 155:75--+, January 1969.

\bibitem{Clayton:1987}
D.~D. {Clayton} and M.~D. {Leising}.
\newblock {$^{26}$Al in the interstellar medium.}
\newblock {\em \physrep}, 144:1--50, 1987.

\bibitem{Clayton:2004}
D.~D. {Clayton} and L.~R. {Nittler}.
\newblock {Astrophysics with Presolar Stardust}.
\newblock {\em \araa}, 42:39--78, September 2004.

\bibitem{Diehl:2002}
R.~{Diehl}.
\newblock {$^{26}$Al production in the Vela and Orion regions}.
\newblock {\em New Astronomy Review}, 46:547--552, July 2002.

\bibitem{Diehl:2007a}
R.~{Diehl}.
\newblock {Nuclear Astrophysics with Gamma-Ray Line Astronomy}.
\newblock In M.~{Busso}, R.~G. {Pizzone}, C.~{Rolfs}, C.~{Spitaleri}, and
  A.~{Tumino}, editors, {\em EAS Publications Series}, volume~27 of {\em EAS
  Publications Series}, pages 83--102, 2007.

\bibitem{Diehl:2013b}
R.~{Diehl}.
\newblock {Cosmic Gamma-Ray Spectroscopy}.
\newblock {\em ArXiv e-prints}, arXiv1307(1307.4198D):1--8, July 2013.

\bibitem{Diehl:2013}
R.~{Diehl}.
\newblock {Nuclear astrophysics lessons from INTEGRAL}.
\newblock {\em Reports on Progress in Physics}, 76(2):026301, February 2013.

\bibitem{Diehl:1995b}
R.~{Diehl}, C.~{Dupraz}, K.~{Bennett}, H.~{Bloemen}, W.~{Hermsen},
  J.~{Knoedlseder}, G.~{Lichti}, D.~{Morris}, J.~{Ryan}, V.~{Schoenfelder},
  H.~{Steinle}, A.~{Strong}, B.~{Swanenburg}, M.~{Varendorff}, and
  C.~{Winkler}.
\newblock {COMPTEL observations of Galactic $^{26}$Al emission.}
\newblock {\em \aap}, 298:445--+, June 1995.

\bibitem{Diehl:2006d}
R.~{Diehl}, H.~{Halloin}, K.~{Kretschmer}, G.~G. {Lichti},
  V.~{Sch{\"o}nfelder}, A.~W. {Strong}, A.~{von Kienlin}, W.~{Wang}, P.~{Jean},
  J.~{Kn{\"o}dlseder}, J.-P. {Roques}, G.~{Weidenspointner}, S.~{Schanne},
  D.~H. {Hartmann}, C.~{Winkler}, and C.~{Wunderer}.
\newblock {Radioactive $^{26}$Al from massive stars in the Galaxy}.
\newblock {\em \nat}, 439:45--47, January 2006.

\bibitem{Diehl:2006c}
R.~{Diehl}, H.~{Halloin}, K.~{Kretschmer}, A.~W. {Strong}, W.~{Wang},
  P.~{Jean}, G.~G. {Lichti}, J.~{Kn{\"o}dlseder}, J.-P. {Roques}, S.~{Schanne},
  V.~{Sch{\"o}nfelder}, A.~{von Kienlin}, G.~{Weidenspointner}, C.~{Winkler},
  and C.~{Wunderer}.
\newblock {$^{26}$Al in the inner Galaxy. Large-scale spectral characteristics
  derived with SPI/INTEGRAL}.
\newblock {\em \aap}, 449:1025--1031, April 2006.

\bibitem{Diehl:2011b}
R.~{Diehl}, D.~H. {Hartmann}, and N.~{Prantzos}.
\newblock {\em {Astronomy with Radioactivities}}, volume 812 of {\em Lecture
  Notes in Physics, Berlin Springer Verlag}.
\newblock Springer: Berlin, Heidelberg, 2011.

\bibitem{Diehl:2003e}
R.~{Diehl}, K.~{Kretschmer}, S.~{Pl{\"u}schke}, M.~{Cervi{\~n}o}, and D.~H.
  {Hartmann}.
\newblock {Gamma-rays from massive stars in Cygnus and Orion}.
\newblock In K.~{van der Hucht}, A.~{Herrero}, and C.~{Esteban}, editors, {\em
  A Massive Star Odyssey: From Main Sequence to Supernova}, volume 212 of {\em
  IAU Symposium}, pages 706--+, 2003.

\bibitem{Diehl:2010}
R.~{Diehl}, M.~G. {Lang}, P.~{Martin}, H.~{Ohlendorf}, T.~{Preibisch},
  R.~{Voss}, P.~{Jean}, {J.-P.} {Roques}, P.~{von Ballmoos}, and W.~{Wang}.
\newblock {Radioactive $^{26}$Al from the Scorpius-Centaurus association}.
\newblock {\em \aap}, 522:A51+, November 2010.

\bibitem{Diehl:2014}
R.~{Diehl}, T.~{Siegert}, W.~{Hillebrandt}, S.~A. {Grebenev}, J.~{Greiner},
  M.~{Krause}, M.~{Kromer}, K.~{Maeda}, F.~{R{\"o}pke}, and S.~{Taubenberger}.
\newblock {Early $^{56}$Ni decay gamma rays from SN2014J suggest an unusual
  explosion}.
\newblock {\em Science}, 345:1162--1165, September 2014.

\bibitem{Diehl:2015}
R.~{Diehl}, T.~{Siegert}, W.~{Hillebrandt}, M.~{Krause}, J.~{Greiner},
  K.~{Maeda}, F.~K. {R{\"o}pke}, S.~A. {Sim}, W.~{Wang}, and X.~{Zhang}.
\newblock {SN2014J gamma rays from the $^{56}$Ni decay chain}.
\newblock {\em \aap}, 574:A72, February 2015.

\bibitem{Done:2007}
C.~{Done}, M.~{Gierli{\'n}ski}, and A.~{Kubota}.
\newblock {Modelling the behaviour of accretion flows in X-ray binaries.
  Everything you always wanted to know about accretion but were afraid to ask}.
\newblock {\em \aapr}, 15:1--66, December 2007.

\bibitem{Fierlinger:2012}
K.~M. {Fierlinger}, A.~{Burkert}, R.~{Diehl}, C.~{Dobbs}, D.~H. {Hartmann},
  M.~{Krause}, E.~{Ntormousi}, and R.~{Voss}.
\newblock {Molecular Cloud Disruption and Chemical Enrichment of the ISM Caused
  by Massive Star Feedback}.
\newblock In R.~{Capuzzo-Dolcetta}, M.~{Limongi}, and A.~{Tornamb{\`e}},
  editors, {\em Advances in Computational Astrophysics: Methods, Tools, and
  Outcome}, volume 453 of {\em Astronomical Society of the Pacific Conference
  Series}, page~25, July 2012.

\bibitem{Fierlinger:2016}
K.~M. {Fierlinger}, A.~{Burkert}, E.~{Ntormousi}, P.~{Fierlinger},
  M.~{Schartmann}, A.~{Ballone}, M.~G.~H. {Krause}, and R.~{Diehl}.
\newblock {Stellar feedback efficiencies: supernovae versus stellar winds}.
\newblock {\em \mnras}, 456:710--730, February 2016.

\bibitem{Fossey:2014aa}
J.~{Fossey}, B.~{Cooke}, G.~{Pollack}, M.~{Wilde}, and T.~{Wright}.
\newblock {Supernova 2014J in M82 = Psn J09554214+6940260}.
\newblock {\em Central Bureau Electronic Telegrams}, 3792:1, January 2014.

\bibitem{Fraternali:2012}
F.~{Fraternali}.
\newblock {Modelling the gas kinematics in disk galaxies}.
\newblock In M.~A. {de Avillez}, editor, {\em EAS Publications Series},
  volume~56 of {\em EAS Publications Series}, pages 355--362, September 2012.

\bibitem{Gehrels:1993}
N.~{Gehrels}, C.~E. {Fichtel}, G.~J. {Fishman}, J.~D. {Kurfess}, and
  V.~{Schonfelder}.
\newblock {The Compton Gamma-Ray Observatory}.
\newblock {\em Scientific American}, 269:68--+, December 1993.

\bibitem{Goobar:2014aa}
A.~{Goobar}, J.~{Johansson}, R.~{Amanullah}, Y.~{Cao}, D.~A. {Perley}, M.~M.
  {Kasliwal}, R.~{Ferretti}, P.~E. {Nugent}, C.~{Harris}, A.~{Gal-Yam}, E.~O.
  {Ofek}, S.~P. {Tendulkar}, M.~{Dennefeld}, S.~{Valenti}, I.~{Arcavi},
  D.~P.~K. {Banerjee}, V.~{Venkataraman}, V.~{Joshi}, N.~M. {Ashok}, S.~B.
  {Cenko}, R.~F. {Diaz}, C.~{Fremling}, A.~{Horesh}, D.~A. {Howell}, S.~R.
  {Kulkarni}, S.~{Papadogiannakis}, T.~{Petrushevska}, D.~{Sand},
  J.~{Sollerman}, V.~{Stanishev}, J.~S. {Bloom}, J.~{Surace}, T.~J. {Dupuy},
  and M.~C. {Liu}.
\newblock {The Rise of SN 2014J in the Nearby Galaxy M82}.
\newblock {\em \apjl}, 784:L12, March 2014.

\bibitem{Grebenev:2012}
S.~A. {Grebenev}, A.~A. {Lutovinov}, S.~{Tsygankov}, and C.~{Winkler}.
\newblock Hard-x-ray emission lines from the decay of $^{44}$ti in the remnant
  of supernova 1987a.
\newblock {\em Nature}, (tbd)((tbd)):(accepted for publication), 2012.

\bibitem{Grefenstette:2014}
B.~W. {Grefenstette}, F.~A. {Harrison}, S.~E. {Boggs}, S.~P. {Reynolds}, C.~L.
  {Fryer}, K.~K. {Madsen}, D.~R. {Wik}, A.~{Zoglauer}, C.~I. {Ellinger}, D.~M.
  {Alexander}, H.~{An}, D.~{Barret}, F.~E. {Christensen}, W.~W. {Craig},
  K.~{Forster}, P.~{Giommi}, C.~J. {Hailey}, A.~{Hornstrup}, V.~M. {Kaspi},
  T.~{Kitaguchi}, J.~E. {Koglin}, P.~H. {Mao}, H.~{Miyasaka}, K.~{Mori},
  M.~{Perri}, M.~J. {Pivovaroff}, S.~{Puccetti}, V.~{Rana}, D.~{Stern}, N.~J.
  {Westergaard}, and W.~W. {Zhang}.
\newblock {Asymmetries in core-collapse supernovae from maps of radioactive
  $^{44}$Ti in CassiopeiaA}.
\newblock {\em \nat}, 506:339--342, February 2014.

\bibitem{Harris:2005}
M.~J. {Harris}, J.~{Kn{\"o}dlseder}, P.~{Jean}, E.~{Cisana}, R.~{Diehl}, G.~G.
  {Lichti}, J.-P. {Roques}, S.~{Schanne}, and G.~{Weidenspointner}.
\newblock {Detection of {$\gamma$}-ray lines from interstellar $^{60}$Fe by the
  high resolution spectrometer SPI}.
\newblock {\em \aap}, 433:L49--L52, April 2005.

\bibitem{Harris:2004}
M.~J. {Harris}, J.~{Knodlseder}, P.~{Jean}, E.~{Cisana}, G.~{Lichti},
  R.~{Diehl}, K.~{Kretschmer}, A.~{Von Kienlin}, J.-P. {Roques}, S.~{Schanne},
  G.~{Weidenspointner}, and C.~{Wunderer}.
\newblock {Preliminary results of INTEGRAL/SPI measurements of radioactive
  $^{60}$Fe in the Galaxy}.
\newblock In {\em Bulletin of the American Astronomical Society}, volume~36 of
  {\em Bulletin of the American Astronomical Society}, pages 950--+, August
  2004.

\bibitem{Harrison:2013}
F.~A. {Harrison}, W.~W. {Craig}, F.~E. {Christensen}, C.~J. {Hailey}, W.~W.
  {Zhang}, S.~E. {Boggs}, D.~{Stern}, W.~R. {Cook}, K.~{Forster}, P.~{Giommi},
  B.~W. {Grefenstette}, Y.~{Kim}, T.~{Kitaguchi}, J.~E. {Koglin}, K.~K.
  {Madsen}, P.~H. {Mao}, H.~{Miyasaka}, K.~{Mori}, M.~{Perri}, M.~J.
  {Pivovaroff}, S.~{Puccetti}, V.~R. {Rana}, N.~J. {Westergaard}, J.~{Willis},
  A.~{Zoglauer}, H.~{An}, M.~{Bachetti}, N.~M. {Barri{\`e}re}, E.~C. {Bellm},
  V.~{Bhalerao}, N.~F. {Brejnholt}, F.~{Fuerst}, C.~C. {Liebe}, C.~B.
  {Markwardt}, M.~{Nynka}, J.~K. {Vogel}, D.~J. {Walton}, D.~R. {Wik}, D.~M.
  {Alexander}, L.~R. {Cominsky}, A.~E. {Hornschemeier}, A.~{Hornstrup}, V.~M.
  {Kaspi}, G.~M. {Madejski}, G.~{Matt}, S.~{Molendi}, D.~M. {Smith}, J.~A.
  {Tomsick}, M.~{Ajello}, D.~R. {Ballantyne}, M.~{Balokovi{\'c}}, D.~{Barret},
  F.~E. {Bauer}, R.~D. {Blandford}, W.~{Niel Brandt}, L.~W. {Brenneman},
  J.~{Chiang}, D.~{Chakrabarty}, J.~{Chenevez}, A.~{Comastri}, F.~{Dufour},
  M.~{Elvis}, A.~C. {Fabian}, D.~{Farrah}, C.~L. {Fryer}, E.~V. {Gotthelf},
  J.~E. {Grindlay}, D.~J. {Helfand}, R.~{Krivonos}, D.~L. {Meier}, J.~M.
  {Miller}, L.~{Natalucci}, P.~{Ogle}, E.~O. {Ofek}, A.~{Ptak}, S.~P.
  {Reynolds}, J.~R. {Rigby}, G.~{Tagliaferri}, S.~E. {Thorsett}, E.~{Treister},
  and C.~M. {Urry}.
\newblock {The Nuclear Spectroscopic Telescope Array (NuSTAR) High-energy X-Ray
  Mission}.
\newblock {\em \apj}, 770:103, June 2013.

\bibitem{Heftrich:2015}
T.~{Heftrich}, M.~{Bichler}, R.~{Dressler}, K.~{Eberhardt}, A.~{Endres},
  J.~{Glorius}, K.~{G{\"o}bel}, G.~{Hampel}, M.~{Heftrich}, F.~{K{\"a}ppeler},
  C.~{Lederer}, M.~{Mikorski}, R.~{Plag}, R.~{Reifarth}, C.~{Stieghorst},
  S.~{Schmidt}, D.~{Schumann}, Z.~{Slavkovsk{\'a}}, K.~{Sonnabend},
  A.~{Wallner}, M.~{Weigand}, N.~{Wiehl}, and S.~{Zauner}.
\newblock {Thermal neutron capture cross section of the radioactive isotope
  $^{60}$Fe}.
\newblock {\em \prc}, 92(1):015806, July 2015.

\bibitem{Hillebrandt:2013aa}
W.~{Hillebrandt}, M.~{Kromer}, F.~K. {R{\"o}pke}, and A.~J. {Ruiter}.
\newblock {Towards an understanding of Type Ia supernovae from a synthesis of
  theory and observations}.
\newblock {\em Frontiers of Physics}, 8:116--143, April 2013.

\bibitem{Iyudin:1994}
A.~F. {Iyudin}, R.~{Diehl}, H.~{Bloemen}, W.~{Hermsen}, G.~G. {Lichti},
  D.~{Morris}, J.~{Ryan}, V.~{Schoenfelder}, H.~{Steinle}, M.~{Varendorff},
  C.~{de Vries}, and C.~{Winkler}.
\newblock {COMPTEL observations of Ti-44 gamma-ray line emission from CAS A}.
\newblock {\em \aap}, 284:L1--L4, April 1994.

\bibitem{Jean:2006}
P.~{Jean}, J.~{Kn{\"o}dlseder}, W.~{Gillard}, N.~{Guessoum}, K.~{Ferri{\`e}re},
  A.~{Marcowith}, V.~{Lonjou}, and J.~P. {Roques}.
\newblock {Spectral analysis of the Galactic positron annihilation emission}.
\newblock {\em \aap}, 445:579--589, January 2006.

\bibitem{Jean:2003a}
P.~{Jean}, J.~{Kn{\"o}dlseder}, V.~{Lonjou}, M.~{Allain}, J.-P. {Roques}, G.~K.
  {Skinner}, B.~J. {Teegarden}, G.~{Vedrenne}, P.~{von Ballmoos}, B.~{Cordier},
  P.~{Caraveo}, R.~{Diehl}, P.~{Durouchoux}, P.~{Mandrou}, J.~{Matteson},
  N.~{Gehrels}, V.~{Sch{\"o}nfelder}, A.~W. {Strong}, P.~{Ubertini},
  G.~{Weidenspointner}, and C.~{Winkler}.
\newblock {Early SPI/INTEGRAL measurements of 511 keV line emission from the
  4th quadrant of the Galaxy}.
\newblock {\em \aap}, 407:L55--L58, August 2003.

\bibitem{Jones:2015}
S.~{Jones}, R.~{Hirschi}, M.~{Pignatari}, A.~{Heger}, C.~{Georgy},
  N.~{Nishimura}, C.~{Fryer}, and F.~{Herwig}.
\newblock {Code dependencies of pre-supernova evolution and nucleosynthesis in
  massive stars: evolution to the end of core helium burning}.
\newblock {\em \mnras}, 447:3115--3129, March 2015.

\bibitem{Knodlseder:2000}
J.~{Kn{\"o}dlseder}.
\newblock {Cygnus OB2 - a young globular cluster in the Milky Way}.
\newblock {\em \aap}, 360:539--548, August 2000.

\bibitem{Knodlseder:2005}
J.~{Kn{\"o}dlseder}, P.~{Jean}, V.~{Lonjou}, G.~{Weidenspointner},
  N.~{Guessoum}, W.~{Gillard}, G.~{Skinner}, P.~{von Ballmoos}, G.~{Vedrenne},
  J.-P. {Roques}, S.~{Schanne}, B.~{Teegarden}, V.~{Sch{\"o}nfelder}, and
  C.~{Winkler}.
\newblock {The all-sky distribution of 511 keV electron-positron annihilation
  emission}.
\newblock {\em \aap}, 441:513--532, October 2005.

\bibitem{Krause:2014a}
M.~{Krause}, R.~{Diehl}, H.~{B{\"o}hringer}, M.~{Freyberg}, and D.~{Lubos}.
\newblock {Feedback by massive stars and the emergence of superbubbles. II.
  X-ray properties}.
\newblock {\em \aap}, 566:A94, June 2014.

\bibitem{Krause:2016}
M.~G.~H. {Krause}, C.~{Charbonnel}, N.~{Bastian}, and R.~{Diehl}.
\newblock {Gas expulsion in massive star clusters?. Constraints from
  observations of young and gas-free objects}.
\newblock {\em \aap}, 587:A53, March 2016.

\bibitem{Krause:2015}
M.~G.~H. {Krause}, R.~{Diehl}, Y.~{Bagetakos}, E.~{Brinks}, A.~{Burkert},
  O.~{Gerhard}, J.~{Greiner}, K.~{Kretschmer}, and T.~{Siegert}.
\newblock {$^{26}$Al kinematics: superbubbles following the spiral arms?.
  Constraints from the statistics of star clusters and HI supershells}.
\newblock {\em \aap}, 578:A113, June 2015.

\bibitem{Kretschmer:2013}
K.~{Kretschmer}, R.~{Diehl}, M.~{Krause}, A.~{Burkert}, K.~{Fierlinger},
  O.~{Gerhard}, J.~{Greiner}, and W.~{Wang}.
\newblock {Kinematics of massive star ejecta in the Milky Way as traced by
  $^{26}$Al}.
\newblock {\em \aap}, 559:A99, November 2013.

\bibitem{Leventhal:1989}
M.~{Leventhal}, C.~J. {MacCallum}, S.~D. {Barthelmy}, N.~{Gehrels}, B.~J.
  {Teegarden}, and J.~{Tueller}.
\newblock {Reappearance of the annihilation line source at the Galactic
  Centre}.
\newblock {\em \nat}, 339:36--38, May 1989.

\bibitem{Limongi:2006qf}
M.~{Limongi} and A.~{Chieffi}.
\newblock {Nucleosynthesis of $^{60}$Fe in massive stars}.
\newblock {\em \nar}, 50:474--476, October 2006.

\bibitem{Limongi:2006}
M.~{Limongi} and A.~{Chieffi}.
\newblock {Nucleosynthesis of $^{60}$Fe in massive stars}.
\newblock {\em New Astronomy Review}, 50:474--476, October 2006.

\bibitem{Lingenfelter:1978}
R.~E. {Lingenfelter} and R.~{Ramaty}.
\newblock {Gamma-ray lines - A new window to the Universe}.
\newblock {\em Physics Today}, 31:40--47, March 1978.

\bibitem{Mahoney:1982}
W.~A. {Mahoney}, J.~C. {Ling}, A.~S. {Jacobson}, and R.~E. {Lingenfelter}.
\newblock {Diffuse galactic gamma-ray line emission from nucleosynthetic Fe-60,
  Al-26, and Na-22 - Preliminary limits from HEAO 3}.
\newblock {\em \apj}, 262:742--+, November 1982.

\bibitem{Martin:2009}
P.~{Martin}, J.~{Kn{\"o}dlseder}, R.~{Diehl}, and G.~{Meynet}.
\newblock {New estimates of the gamma-ray line emission of the Cygnus region
  from INTEGRAL/SPI observations}.
\newblock {\em \aap}, 506:703--710, November 2009.

\bibitem{Martin:2010b}
P.~{Martin}, J.~{Kn{\"o}dlseder}, G.~{Meynet}, and R.~{Diehl}.
\newblock {Predicted gamma-ray line emission from the Cygnus complex}.
\newblock {\em \aap}, 511:A86+, February 2010.

\bibitem{Martin:2012}
P.~{Martin}, A.~W. {Strong}, P.~{Jean}, A.~{Alexis}, and R.~{Diehl}.
\newblock {Galactic annihilation emission from nucleosynthesis positrons}.
\newblock {\em \aap}, 543:A3, July 2012.

\bibitem{Mirabel:1999}
I.~F. {Mirabel} and L.~F. {Rodr{\'{\i}}guez}.
\newblock {Sources of Relativistic Jets in the Galaxy}.
\newblock {\em \araa}, 37:409--443, 1999.

\bibitem{Mirabel:1992}
I.~F. {Mirabel}, L.~F. {Rodriguez}, B.~{Cordier}, J.~{Paul}, and F.~{Lebrun}.
\newblock {A double-sided radio jet from the compact Galactic Centre
  annihilator 1E140.7 - 2942}.
\newblock {\em \nat}, 358:215--217, July 1992.

\bibitem{Oberlack:1996}
U.~{Oberlack}, K.~{Bennett}, H.~{Bloemen}, R.~{Diehl}, C.~{Dupraz},
  W.~{Hermsen}, J.~{Knoedlseder}, D.~{Morris}, V.~{Schoenfelder}, A.~{Strong},
  and C.~{Winkler}.
\newblock {The COMPTEL 1.809MeV all-sky image.}
\newblock {\em \aaps}, 120:C311, December 1996.

\bibitem{Pignatari:2010}
M.~{Pignatari}, R.~{Gallino}, M.~{Heil}, M.~{Wiescher}, F.~{K{\"a}ppeler},
  F.~{Herwig}, and S.~{Bisterzo}.
\newblock {The Weak s-Process in Massive Stars and its Dependence on the
  Neutron Capture Cross Sections}.
\newblock {\em \apj}, 710:1557--1577, February 2010.

\bibitem{Pignatari:2013}
M.~{Pignatari}, R.~{Hirschi}, M.~{Wiescher}, R.~{Gallino}, M.~{Bennett},
  M.~{Beard}, C.~{Fryer}, F.~{Herwig}, G.~{Rockefeller}, and F.~X. {Timmes}.
\newblock {The $^{12}$C + $^{12}$C Reaction and the Impact on Nucleosynthesis
  in Massive Stars}.
\newblock {\em \apj}, 762:31, January 2013.

\bibitem{Pignatari:2013a}
M.~{Pignatari}, M.~{Wiescher}, F.~X. {Timmes}, R.~J. {de Boer}, F.-K.
  {Thielemann}, C.~{Fryer}, A.~{Heger}, F.~{Herwig}, and R.~{Hirschi}.
\newblock {Production of Carbon-rich Presolar Grains from Massive Stars}.
\newblock {\em \apjl}, 767:L22, April 2013.

\bibitem{Pluschke:2001c}
S.~{Pl{\"u}schke}, R.~{Diehl}, V.~{Sch{\"o}nfelder}, H.~{Bloemen},
  W.~{Hermsen}, K.~{Bennett}, C.~{Winkler}, M.~{McConnell}, J.~{Ryan},
  U.~{Oberlack}, and J.~{Kn{\"o}dlseder}.
\newblock {The COMPTEL 1.809 MeV survey}.
\newblock In A.~{Gimenez}, V.~{Reglero}, and C.~{Winkler}, editors, {\em
  Exploring the Gamma-Ray Universe}, volume 459 of {\em ESA Special
  Publication}, pages 55--58, September 2001.

\bibitem{Poppel:2010}
W.~G.~L. {P{\"o}ppel}, E.~{Bajaja}, E.~M. {Arnal}, and R.~{Morras}.
\newblock {The interstellar medium surrounding the Scorpius-Centaurus
  association revisited}.
\newblock {\em \aap}, 512:A83+, March 2010.

\bibitem{Prantzos:2011}
N.~{Prantzos}, C.~{Boehm}, A.~M. {Bykov}, R.~{Diehl}, K.~{Ferri{\`e}re},
  N.~{Guessoum}, P.~{Jean}, J.~{Knoedlseder}, A.~{Marcowith}, I.~V.
  {Moskalenko}, A.~{Strong}, and G.~{Weidenspointner}.
\newblock {The 511 keV emission from positron annihilation in the Galaxy}.
\newblock {\em Reviews of Modern Physics}, 83:1001--1056, July 2011.

\bibitem{Prantzos:1996a}
N.~{Prantzos} and R.~{Diehl}.
\newblock {Radioactive 26Al in the galaxy: observations versus theory}.
\newblock {\em \physrep}, 267:1--69, March 1996.

\bibitem{Preibisch:2002}
T.~{Preibisch}, A.~G.~A. {Brown}, T.~{Bridges}, E.~{Guenther}, and
  H.~{Zinnecker}.
\newblock {Exploring the Full Stellar Population of the Upper Scorpius OB
  Association}.
\newblock {\em \aj}, 124:404--416, July 2002.

\bibitem{Purcell:1997}
W.~R. {Purcell}, L.-X. {Cheng}, D.~D. {Dixon}, R.~L. {Kinzer}, J.~D. {Kurfess},
  M.~{Leventhal}, M.~A. {Saunders}, J.~G. {Skibo}, D.~M. {Smith}, and
  J.~{Tueller}.
\newblock {OSSE Mapping of Galactic 511 keV Positron Annihilation Line
  Emission}.
\newblock {\em \apj}, 491:725--+, December 1997.

\bibitem{Ramaty:1992}
R.~{Ramaty}, M.~{Leventhal}, K.~W. {Chan}, and R.~E. {Lingenfelter}.
\newblock {On the origin of variable 511 keV line emission from the Galactic
  center region}.
\newblock {\em \apjl}, 392:L63--L66, June 1992.

\bibitem{Ramaty:1994}
R.~{Ramaty}, J.~G. {Skibo}, and R.~E. {Lingenfelter}.
\newblock {Galactic 0.511 MeV line emission}.
\newblock {\em \apjs}, 92:393--399, June 1994.

\bibitem{Remillard:2006}
R.~A. {Remillard} and J.~E. {McClintock}.
\newblock {X-Ray Properties of Black-Hole Binaries}.
\newblock {\em \araa}, 44:49--92, September 2006.

\bibitem{Rodriguez:2015}
J.~{Rodriguez}, M.~{Cadolle Bel}, J.~{Alfonso-Garz{\'o}n}, T.~{Siegert}, X.-L.
  {Zhang}, V.~{Grinberg}, V.~{Savchenko}, J.~A. {Tomsick}, J.~{Chenevez},
  M.~{Clavel}, S.~{Corbel}, R.~{Diehl}, A.~{Domingo}, C.~{Gouiff{\`e}s},
  J.~{Greiner}, M.~G.~H. {Krause}, P.~{Laurent}, A.~{Loh}, S.~{Markoff}, J.~M.
  {Mas-Hesse}, J.~C.~A. {Miller-Jones}, D.~M. {Russell}, and J.~{Wilms}.
\newblock {Correlated optical, X-ray, and {$\gamma$}-ray flaring activity seen
  with INTEGRAL during the 2015 outburst of V404 Cygni}.
\newblock {\em \aap}, 581:L9, September 2015.

\bibitem{Ropke:2011aa}
F.~K. {R{\"o}pke}, I.~R. {Seitenzahl}, S.~{Benitez}, M.~{Fink}, R.~{Pakmor},
  M.~{Kromer}, S.~A. {Sim}, F.~{Ciaraldi-Schoolmann}, and W.~{Hillebrandt}.
\newblock {Modeling Type Ia supernova explosions}.
\newblock {\em Progress in Particle and Nuclear Physics}, 66:309--318, April
  2011.

\bibitem{Rugel:2009}
G.~{Rugel}, T.~{Faestermann}, K.~{Knie}, G.~{Korschinek}, M.~{Poutivtsev},
  D.~{Schumann}, N.~{Kivel}, I.~{G{\"u}nther-Leopold}, R.~{Weinreich}, and
  M.~{Wohlmuther}.
\newblock {New Measurement of the Fe60 Half-Life}.
\newblock {\em Physical Review Letters}, 103(7):072502--+, August 2009.

\bibitem{Schoenfelder:1996}
V.~{Schoenfelder}, K.~{Bennett}, H.~{Bloemen}, R.~{Diehl}, W.~{Hermsen},
  G.~{Lichti}, M.~{McConnell}, J.~{Ryan}, A.~{Strong}, and C.~{Winkler}.
\newblock {COMPTEL overview: Achievements and expectations.}
\newblock {\em \aaps}, 120:C13+, November 1996.

\bibitem{Share:1985}
G.~H. {Share}, R.~L. {Kinzer}, J.~D. {Kurfess}, D.~J. {Forrest}, E.~L. {Chupp},
  and E.~{Rieger}.
\newblock {Detection of galactic Al-26 gamma radiation by the SMM
  spectrometer}.
\newblock {\em \apjl}, 292:L61--L65, May 1985.

\bibitem{Siegert:2016a}
R.; {Greiner} J.; et~al. {Siegert}, T.;~{Diehl}.
\newblock Positron annihilation signatures associated with teh outburst of the
  microquasar v404 cygni.
\newblock {\em Nature}, 531(7592):?, March 2016.

\bibitem{Siegert:2016}
T.~{Siegert}, R.~{Diehl}, G.~{Khachatryan}, M.~G.~H. {Krause},
  F.~{Guglielmetti}, J.~{Greiner}, A.~W. {Strong}, and X.~{Zhang}.
\newblock {Gamma-ray spectroscopy of positron annihilation in the Milky Way}.
\newblock {\em \aap}, 586:A84, February 2016.

\bibitem{Siegert:2015}
T.~{Siegert}, R.~{Diehl}, M.~G.~H. {Krause}, and J.~{Greiner}.
\newblock {Revisiting INTEGRAL/SPI observations of $^{44}$Ti from Cassiopeia
  A}.
\newblock {\em \aap}, 579:A124, July 2015.

\bibitem{Signore:1993}
M.~{Signore} and C.~{Dupraz}.
\newblock {Massive stars as Galactic producers of Al-26}.
\newblock {\em \aaps}, 97:141--144, January 1993.

\bibitem{Skinner:2012}
G.~{Skinner}, P.~{Jean}, J.~{Knoedlseder}, P.~{von Ballmoos}, M.~{Leising},
  P.~{Milne}, and G.~{Weidenspointner}.
\newblock {The 511 keV sky as seen by INTEGRAL/SPI, CGRO-OSSE and GRS/SMM
  combined}.
\newblock In {\em Proceedings of ''An INTEGRAL view of the high-energy sky (the
  first 10 years)'' - 9th INTEGRAL Workshop and celebration of the 10th
  anniversary of the launch (INTEGRAL 2012). 15-19 October 2012. Bibliotheque
  Nationale de France, Paris, France. Published online at <A
  href=''http://pos.sissa.it/cgi-bin/reader/conf.cgi?confid=176''>http://pos.sissa.it/cgi-bin/reader/conf.cgi?confid=176</A>,
  id.112}, page 112, 2012.

\bibitem{The:2014}
L.-S. {The} and A.~{Burrows}.
\newblock {Expectations for the Hard X-Ray Continuum and Gamma-Ray Line Fluxes
  from the Type Ia Supernova SN 2014J in M82}.
\newblock {\em \apj}, 786:141, May 2014.

\bibitem{Timmes:1995}
F.~X. {Timmes}, S.~E. {Woosley}, D.~H. {Hartmann}, R.~D. {Hoffman}, T.~A.
  {Weaver}, and F.~{Matteucci}.
\newblock {26Al and 60Fe from Supernova Explosions}.
\newblock {\em \apj}, 449:204--+, August 1995.

\bibitem{Trombka:1978}
J.~{Trombka}, C.~{Fichtel}, J.~{Grindlay}, and R.~{Hofstadter}.
\newblock {Gamma-ray astrophysics - A new look at the universe}.
\newblock {\em Science}, 202:933--938, December 1978.

\bibitem{Trombka:1983}
J.~I. {Trombka} and C.~E. {Fichtel}.
\newblock {Gamma-ray astrophysics.}
\newblock {\em \physrep}, 97:173--218, 1983.

\bibitem{Vedrenne:2003}
G.~{Vedrenne}, J.-P. {Roques}, V.~{Sch{\"o}nfelder}, P.~{Mandrou}, G.~G.
  {Lichti}, A.~{von Kienlin}, B.~{Cordier}, S.~{Schanne}, J.~{Kn{\"o}dlseder},
  G.~{Skinner}, P.~{Jean}, F.~{Sanchez}, P.~{Caraveo}, B.~{Teegarden}, P.~{von
  Ballmoos}, L.~{Bouchet}, P.~{Paul}, J.~{Matteson}, S.~{Boggs}, C.~{Wunderer},
  P.~{Leleux}, G.~{Weidenspointner}, P.~{Durouchoux}, R.~{Diehl}, A.~{Strong},
  M.~{Cass{\'e}}, M.~A. {Clair}, and Y.~{Andr{\'e}}.
\newblock {SPI: The spectrometer aboard INTEGRAL}.
\newblock {\em \aap}, 411:L63--L70, November 2003.

\bibitem{Vink:2004}
J.~{Vink}.
\newblock {X- and {$\gamma$}-ray studies of Cas A: exposing core collapse to
  the core}.
\newblock {\em \nar}, 48:61--67, February 2004.

\bibitem{Voss:2009}
R.~{Voss}, R.~{Diehl}, D.~H. {Hartmann}, M.~{Cervi{\~n}o}, J.~S. {Vink},
  G.~{Meynet}, M.~{Limongi}, and A.~{Chieffi}.
\newblock {Using population synthesis of massive stars to study the
  interstellar medium near OB associations}.
\newblock {\em \aap}, 504:531--542, September 2009.

\bibitem{Voss:2010a}
R.~{Voss}, R.~{Diehl}, J.~S. {Vink}, and D.~H. {Hartmann}.
\newblock {Probing the evolving massive star population in Orion with kinematic
  and radioactive tracers}.
\newblock {\em \aap}, 520:A51+, September 2010.

\bibitem{Voss:2012}
R.~{Voss}, P.~{Martin}, R.~{Diehl}, J.~S. {Vink}, D.~H. {Hartmann}, and
  T.~{Preibisch}.
\newblock {Energetic feedback and $^{26}$Al from massive stars and their
  supernovae in the Carina region}.
\newblock {\em \aap}, 539:A66, March 2012.

\bibitem{Wang:2007a}
W.~{Wang}, M.~J. {Harris}, R.~{Diehl}, H.~{Halloin}, B.~{Cordier}, A.~W.
  {Strong}, K.~{Kretschmer}, J.~{Kn{\"o}dlseder}, P.~{Jean}, G.~G. {Lichti},
  J.~P. {Roques}, S.~{Schanne}, A.~{von Kienlin}, G.~{Weidenspointner}, and
  C.~{Wunderer}.
\newblock {SPI observations of the diffuse $^{60}$Fe emission in the Galaxy}.
\newblock {\em \aap}, 469:1005--1012, July 2007.

\bibitem{Weidenspointner:2008a}
G.~{Weidenspointner}, G.~{Skinner}, P.~{Jean}, J.~{Kn{\"o}dlseder}, P.~{von
  Ballmoos}, G.~{Bignami}, R.~{Diehl}, A.~W. {Strong}, B.~{Cordier},
  S.~{Schanne}, and C.~{Winkler}.
\newblock {An asymmetric distribution of positrons in the Galactic disk
  revealed by {$\gamma$}-rays}.
\newblock {\em \nat}, 451:159--162, January 2008.

\bibitem{Weidner:2005}
C.~{Weidner} and P.~{Kroupa}.
\newblock {The Variation of Integrated Star Initial Mass Functions among
  Galaxies}.
\newblock {\em \apj}, 625:754--762, June 2005.

\bibitem{Weidner:2010}
C.~{Weidner}, P.~{Kroupa}, and I.~A.~D. {Bonnell}.
\newblock {The relation between the most-massive star and its parental star
  cluster mass}.
\newblock {\em \mnras}, 401:275--293, January 2010.

\bibitem{Winkler:2003}
C.~{Winkler}, T.~J.-L. {Courvoisier}, G.~{Di Cocco}, N.~{Gehrels},
  A.~{Gim{\'e}nez}, S.~{Grebenev}, W.~{Hermsen}, J.~M. {Mas-Hesse},
  F.~{Lebrun}, N.~{Lund}, G.~G.~C. {Palumbo}, J.~{Paul}, J.-P. {Roques},
  H.~{Schnopper}, V.~{Sch{\"o}nfelder}, R.~{Sunyaev}, B.~{Teegarden},
  P.~{Ubertini}, G.~{Vedrenne}, and A.~J. {Dean}.
\newblock {The INTEGRAL mission}.
\newblock {\em \aap}, 411:L1--L6, November 2003.

\end{thebibliography}
%\end{thebibliography}

\end{document}